\documentclass[fleqn,usenatbib]{mnras}

\usepackage[T1]{fontenc}
\usepackage{ae,aecompl}
\usepackage{hyperref}
\usepackage{bm}

\usepackage[dvips]{graphicx}	
\usepackage{amsmath}	
\usepackage{amssymb} 
\usepackage{footnote}
\usepackage{adjustbox}
\usepackage{lipsum}
\usepackage{mwe}
\usepackage{verbatim}
\usepackage{blindtext}
\usepackage{mathtools}
\usepackage{relsize}
\usepackage[group-separator={,},group-minimum-digits=4]{siunitx}
\usepackage{subfig,float}
\usepackage{rotating}
\usepackage{adjustbox}
\usepackage{multirow}

\newcommand{\Msun}{${\rm M_\odot}$}
\newcommand{\Lsun}{${\rm L_\odot}$}
\newcommand{\Rsun}{${\rm R_\odot}$}

\newcommand{\Teff}{${\rm T_{eff}}$}
\newcommand{\Mhb}{${\rm M_{HB}}$}
\newcommand{\Lbol}{${\rm L_{bol}}$}

\title[UVIT study of M3 and M13]{Globular Cluster UVIT Legacy Survey (GlobULeS) $-$ II. Evolutionary status of hot stars in M3 and M13}

\author[Kumar et al.]{
Ranjan Kumar $^{1}$\thanks{E-mail: ranjankmr488@gmail.com}, 
Ananta C. Pradhan $^{1}$\thanks{E-mail: acp.phy@gmail.com},
Snehalata Sahu $^{2}$,
\newauthor Annapurni Subramaniam $^{3}$,
Sonika Piridi $^{1}$,
 Santi Cassisi $^{4,5}$,
Devendra K. Ojha $^{6}$ \\
$^{1}$Department of Physics and Astronomy, National Institute of Technology, Rourkela, Odisha - 769 008, India\\
$^{2}$Department of Physics, University of Warwick, Coventry, CV4 7AL, UK\\
$^{3}$Indian Institute of Astrophysics, Koramangala II Block, Bangalore-560034, India\\
$^{4}$INAF - Astronomical Observatory of Abruzzo, Via M. Maggini, sn. 64100 Teramo, Italy.\\
$^{5}$INFN -  Sezione di Pisa, Universit\`a di Pisa, Largo Pontecorvo 3, 56127 Pisa, Italy.\\
$^{6}$Department of Astronomy and Astrophysics, Tata Institute of Fundamental Research (TIFR), Mumbai - 400 005, India\\
}

\date{Accepted 2023 March 30. Received 2023 March 16; in original form 2022 September 3}

\pubyear{2023}

\begin{document}
\label{firstpage}
\pagerange{\pageref{firstpage}--\pageref{lastpage}}
\maketitle

\begin{abstract}
We present a far-ultraviolet (FUV) study of hot stellar populations in the second parameter pair globular clusters (GCs) M3 and M13, as a part of the Globular cluster UVIT Legacy Survey program (GlobULeS). We use observations made with F148W and F169M filters of the Ultraviolet Imaging Telescope (UVIT) onboard {\em AstroSat} along with ground-based data (UBVRI filters), {\em Hubble Space Telescope (HST)} GC catalogue, and {\em Gaia} EDR3 catalogue. Based on the FUV-optical colour-magnitude diagrams, we classify the sources into  the horizontal branch (HB) stars, post-HB stars, and hot white dwarfs (WDs) in both the GCs. 
The comparison of synthetic and observed colours of the observed HB stars suggests that the mass-loss at the red giant branch (RGB) and He spread in both clusters have a simultaneous effect on the different HB distributions detected in M3 and M13, such that, HB stars of M13 require a larger spread in He (${\rm 0.247-0.310}$) than those of M3 (${\rm Y= 0.252-0.266}$). The evolutionary status of HB stars, post-HB stars, and WDs are studied using SED fit parameters and theoretical evolutionary tracks on the H-R diagram. We found that the observed post-HB stars have evolved from zero-age HB (ZAHB) stars of the mass range $0.48-0.55$ \Msun\ in M3 and M13. We detect 24 WD candidates in each cluster having ${\rm \log(L_{bol}/L_\odot)}$ in the range $-0.8$ to $+0.6$ and ${\rm \log(T_{eff}/K)}$ in the range of 4.2 to 5.0. Placing the WDs on the H-R diagram and comparing them with models suggest that M13 has a population of low-mass WDs, probably originating from binary evolution.  

\end{abstract}
\begin{keywords}
stars: horizontal branch$-$white dwarf$-$Hertzsprung-Russell and colour-magnitude diagrams$-$ (Galaxy:) globular clusters: individual: NGC 5272 (M3) and NGC 6205 (M13) $-$ ultraviolet: stars 
\end{keywords}

\section{Introduction}   \label{sec:introduction}

M3 (NGC $5272$) and M13 (NGC $6205$) are regarded as two twin globular clusters (GCs) in many aspects. They have been studied extensively photometrically, particularly, in optical and ultraviolet (UV) wavebands to explore the peculiar features of their horizontal branch (HB) stars \citep{Newell1970, Peterson1983, Laget1992, Catelan1995, Whitney1995, Ferraro1997, Behr2003, Moehler2003, Dalessandro2013, Denissenkov2017, Chen2021}. Despite having similar metallicity (${\rm [Fe/H]}\sim-1.5$ dex) and age ($\sim12.5$ Gyr) \citep{Harris1996}, their HB morphologies are different. In optical colour-magnitude diagrams (CMDs), HB stars of M3 show a horizontal sequence, covering red HB, RR Lyrae (RRL), and blue HB \citep{Ferraro1997}, whereas HB stars of M13 show a vertical sequence, covering extreme HB, blue HB, and a very small number of RRL variables \citep{Paltrinieri1998}. The morphology of HB stars in M3 and M13 has served as the best suitable example of the long-standing ``second parameter problem'' for decades \citep{Sandage1967, Rood1973, Catelan1995, Moehler2003, Dalessandro2013}. Several investigations were carried out to identify the second parameter that governs the HB morphology in Galactic GCs, but not a single parameter is confirmed yet \citep[for a detailed review see][ and references therein]{Moehler2010, Dotter2013, Milone2014, Cassisi2020, Milone2022}.  

Multiple populations (MPs) in GC play a vital role in placing the core He-burning stars of the cluster on the HB. Second-generation (2G) stars are usually He-enhanced \citep{Milone2018}. They evolve more quickly and lose more mass along the RGB phase than the first-generation (1G) stars which are typically He-normal. As a result, 2G stars on HB have higher effective temperature (\Teff) than the 1G stars \citep{Tailo2019, Tailo2020}. Both the extreme HB stars of M13 and the blue HB stars of M3 are the progeny of 2G RGB stars that are enriched with He and have had considerably greater mass-loss along the RGB phase \citep{Tailo2020}. With the aid of the multi-band HST globular cluster survey data, there have been numerous advancements made over the past ten years in the study of multiple populations in GCs \citep[][and references therein]{Milone2022}. However, their impact on shaping the HB morphology in GCs has been mostly contributed by the variation in He-abundances \citep{Marino2014, Brown2016, Milone2018} and the mass-loss along the RGB phase of the 1G and 2G populations \citep{Tailo2019, Tailo2020}. 

In the last decade, the observed HB morphology in M3 and M13 was studied by \cite{Dalessandro2013, Denissenkov2017, Tailo2019} and \cite{Tailo2020} using He-spread and mass-loss prescriptions. They have all stated that the observed HB sequence in M13 requires a relatively higher He-spread as well as mass-loss  than that in M3. In order to produce the observed HB sequence of M3, a similar mass-loss and He-spread is required. In order to produce the observed HB sequence of M13, \cite{Dalessandro2013} reported a He-spread (${\rm \Delta Y}$) of 0.05 dex with maximum mass-loss along RGB (${\rm \Delta M}$)  of 0.266 \Msun, \cite{Denissenkov2017} reported ${\rm \Delta Y}$ of 0.08 dex with maximum ${\rm \Delta M}$ of 0.2165 \Msun, and \cite{Tailo2020} reported maximum ${\rm \Delta M}$ of 0.273 $\pm$ 0.021 \Msun\ using ${\rm \Delta Y}$ of 0.052 $\pm$ 0.004 dex derived by \cite{Milone2018}. The above-reported values suggest that the estimates of He-spread and mass-loss by \cite{Dalessandro2013} are still valid within the uncertainties of the values derived by \cite{Milone2018} and \cite{Tailo2020} using the latest information on multiple populations in GCs. However, further investigation is certainly required in order to find the best relation between He-spread and mass-loss along RGB in order to produce the HB morphology of Galactic GCs. 

Stars in the post-HB phase evolve faster than their previous evolutionary phases, hence, we see a very small number of such stars in GCs. Based on their morphology in the HB phase, the progeny of red HB, blue HB, and extreme HB stars evolve differently. A detailed analysis on the post-HB evolution of stars in GCs can be found in \citet{Moehler2019}, \citet{Bond2021}, and \citet{Davis2021}. The number and nature of post-HB stars in M3 and M13 are different due to their dissimilar morphology in the HB phase. In the case of M3, HB stars are mostly dominated by red HB to cooler blue HB stars whereas post-HB stars are mostly dominated by AGB stars \citep{Chen2021}. Seven AGB-manqu\'e stars were detected in M13 by \citet{Chen2021} and also they found a relatively smaller fraction of AGB/HB ratio  in M13 ($0.08\pm0.01$) than in M3 ($0.13\pm0.02$). However, there is no proper statistics on the AGB-manqu\'e or hotter post-HB stars yet in these two clusters. This issue might be resolved with UV photometry as in UV they are relatively brighter and easily identifiable than in optical photometry.

The different stellar evolutionary scenarios of the progenitors are likely to affect their white dwarf (WD) population as well. By analysing the high-resolution photometric data of {\em HST}, \citet{Chen2021} found an overabundance ($1.5$ times) of bright WDs in M13  in comparison to M3. From their theoretical models, they argued that the observed overabundance of WDs in M13 might be due to an excess of extreme HB stars in M13 and subsequently a larger fraction of post-HB stars which do not reach the AGB phase (i.e., PEAGB/AGB-manqu\'e stars which do not experience the third dredge-up). Such stars might retain an excess amount of hydrogen on their surface which is burned through a stable nuclear reaction during their WD cooling phase \citep{Miller2013, Althaus2015}. This excess hydrogen burning slows down the WD cooling process which delays the cooling time of such stars and hence we observe a larger fraction of WDs at their hotter/brighter end of the cooling sequence. An UV analysis of WDs in M3 and M13 will certainly benefit to test this theory.
 
The Ultraviolet Imaging Telescope (UVIT) onboard {\em AstroSat} has been playing an important role in exploring the hot HB stars in UV in several GCs. Using UVIT filters, \citet{Kumar2022} detected two new hot sources in M68; one extreme HB star and a post-blue hook star. \citet{Kumar2021JApA, Kumar2021MNRAS} studied the hot blue HB populations in two, GCs NGC 4147 and NGC 7492, and detected one new extreme HB star in NGC 7492. \citet{Prabhu2021} studied the physical properties of post-HB stars in NGC $2808$ and segregated them into different evolutionary phases such as AGB-manqu\'e, PEAGB, post-AGB, etc. Hot stellar populations such as extreme HB stars, hot WD binaries, blue stragglers, etc., have been studied in two GCs, NGC 288 and NGC 1851 by \citet{Sahu2019} and \citet{Singh2020}, respectively using UVIT photometry. Recently, \citet{sneha2020} have made a comprehensive study of UV bright sources in 11 GCs and prepared a photometric catalogue of the sources up to 23.5 mag in F148W filter of UVIT. They were able to detect Grundhal jump \citep[G-jump;][]{Grundahl1998} and Momany jump \citep[M-jump;][]{Momany2004} corresponding to \Teff\ 11,500 K and 21,000 K in HB stars of M13 using pseudo colour (${\rm F148W_{UVIT} - 2 \times F336W_{HST} + F606W_{HST}}$) versus colour (${\rm F148W_{UVIT} - F606W_{HST} }$) plot. In this paper, we aim to perform a comprehensive study of the HB morphology of M3 and M13 GCs and explore their further evolutionary status, such as AGB-manqu\'e, PEAGB, post-AGB and hence, the final product, WDs, in both the clusters.
 
We present the data reduction and photometry procedures of the UVIT observation of M3 and M13 in \S\ref{sec:observation}. In \S\ref{sec:cmd}, we discuss the UV-optical CMDs and the classification of sources. In \S\ref{sec:synth_hb}, a validation of He-spread on the HB of M3 and M13 is performed by comparing the observed HB stars with synthetic HB stars. In \S\ref{sec:post_hb}, we have explored the evolutionary status of the post-HB stars. In \S\ref{sec:wd}, we have studied the spectral energy distribution (SED) of WD candidates and discussed their evolutionary scenario. Finally, we have concluded our results in \S\ref{sec:conclusion}.

\section{Observation and Data Reduction} \label{sec:observation}

We observed M3 and M13 in two FUV filters, F148W and F169M of UVIT. The exposure times of M3 (M13) are $3,000$ ($6,654$) and $2,883$ ($6,657$) seconds in F148W and F169M filters, respectively. The observation, photometry, and cluster membership details of the observed sources are provided in \cite{sneha2020}. The total number of sources detected in M3 and M13 are 878 and 1090, respectively. The completeness of the observed sources in M3 and M13 is above 90\% each outside the half-light radius (${\rm r_h}$) of the clusters, and it is 80 - 90\% and 60 - 70\%, respectively, for sources in between core-radius (${\rm r_c}$) and ${\rm r_h}$ of the clusters, and $\sim$50\% each for sources within ${\rm r_c}$, respectively. The sources present in the inner region i.e., within $3.5'\times3.5'$ of the cluster core were cross-matched with {\em HST} GC catalogue \citep{Nardiello2018} and the rest of the sources beyond this core region were cross-matched with {\em Gaia} EDR3 catalogue \citep{Vasiliev2021}. We retain only those sources  which have a cluster membership probability of more than $90\%$ in the catalogue. Finally, we were left with $274$ ($399$) stars in the inner region ({\em HST} counterparts) and $160$ ($390$) stars in the outer region ({\em Gaia} counterparts) of the cluster M3 (M13). The {\em Gaia} counterparts were then cross-matched with the ground-based photometric catalogue to obtain the magnitudes in U, B, V, R, and I filters \citep{Stetson2019}.
All the observed sources were extinction corrected using the extinction values (${\rm E(B-V)}=0.011$ mag for M3 and $0.014$ mag for M13) obtained from dust extinction map of \citet{Schlafly2011} and extinction law of \citet{cardeli1989}.

\section{colour-Magnitude Diagrams} \label{sec:cmd}

\begin{figure*}
    \centering
    \includegraphics[width=0.49\textwidth]{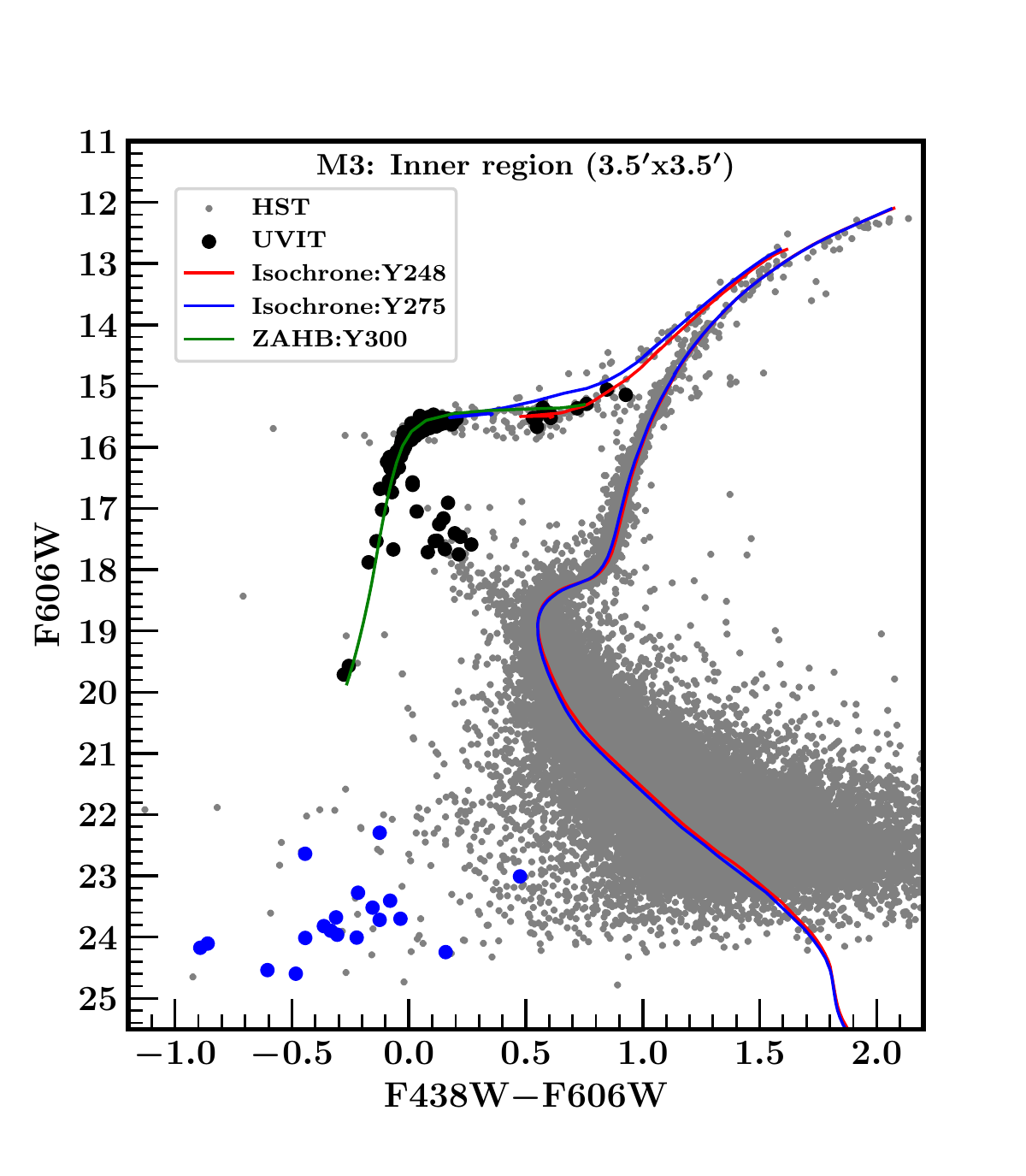}
    \includegraphics[width=0.49\textwidth]{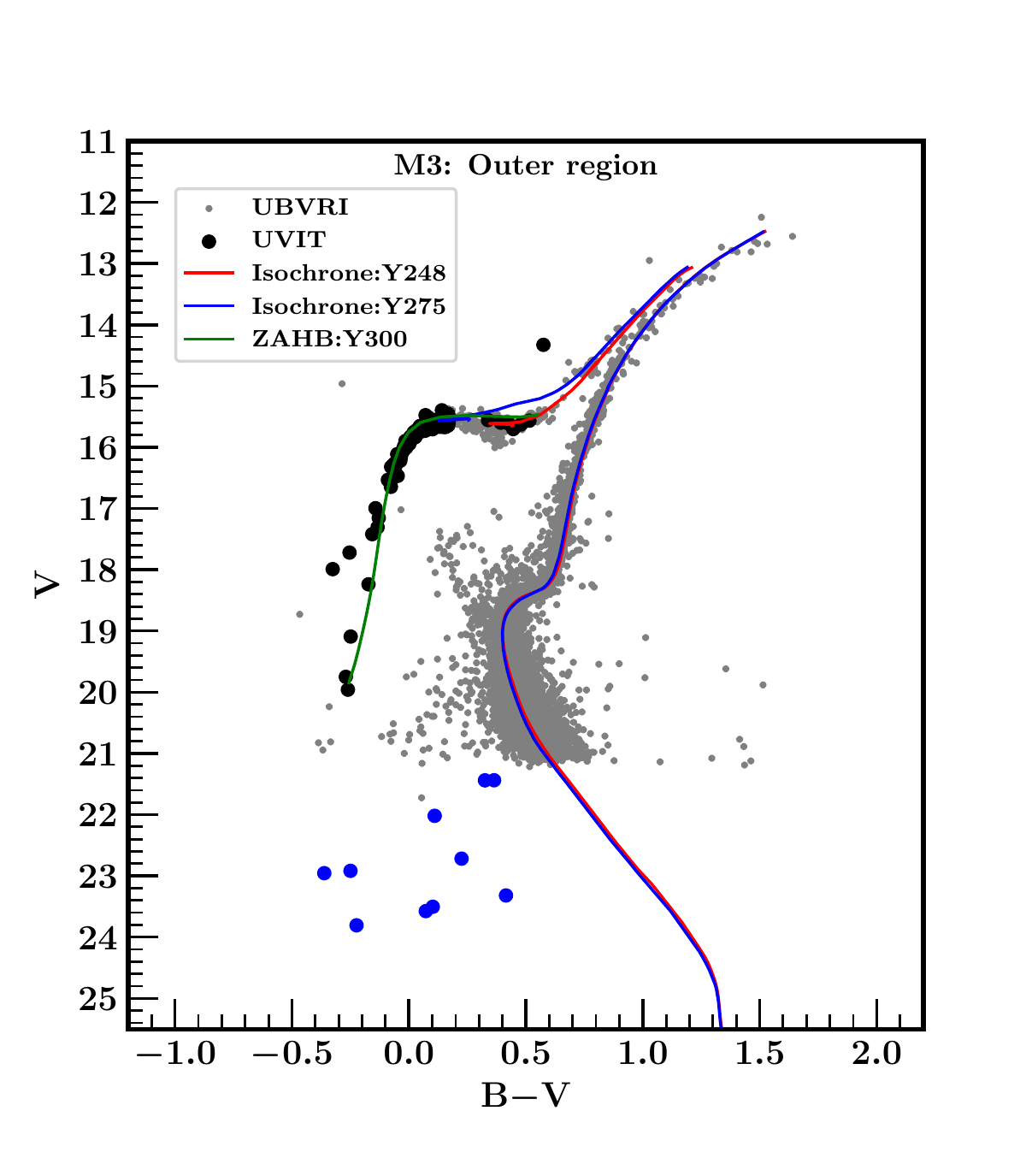}
    \includegraphics[width=0.49\textwidth]{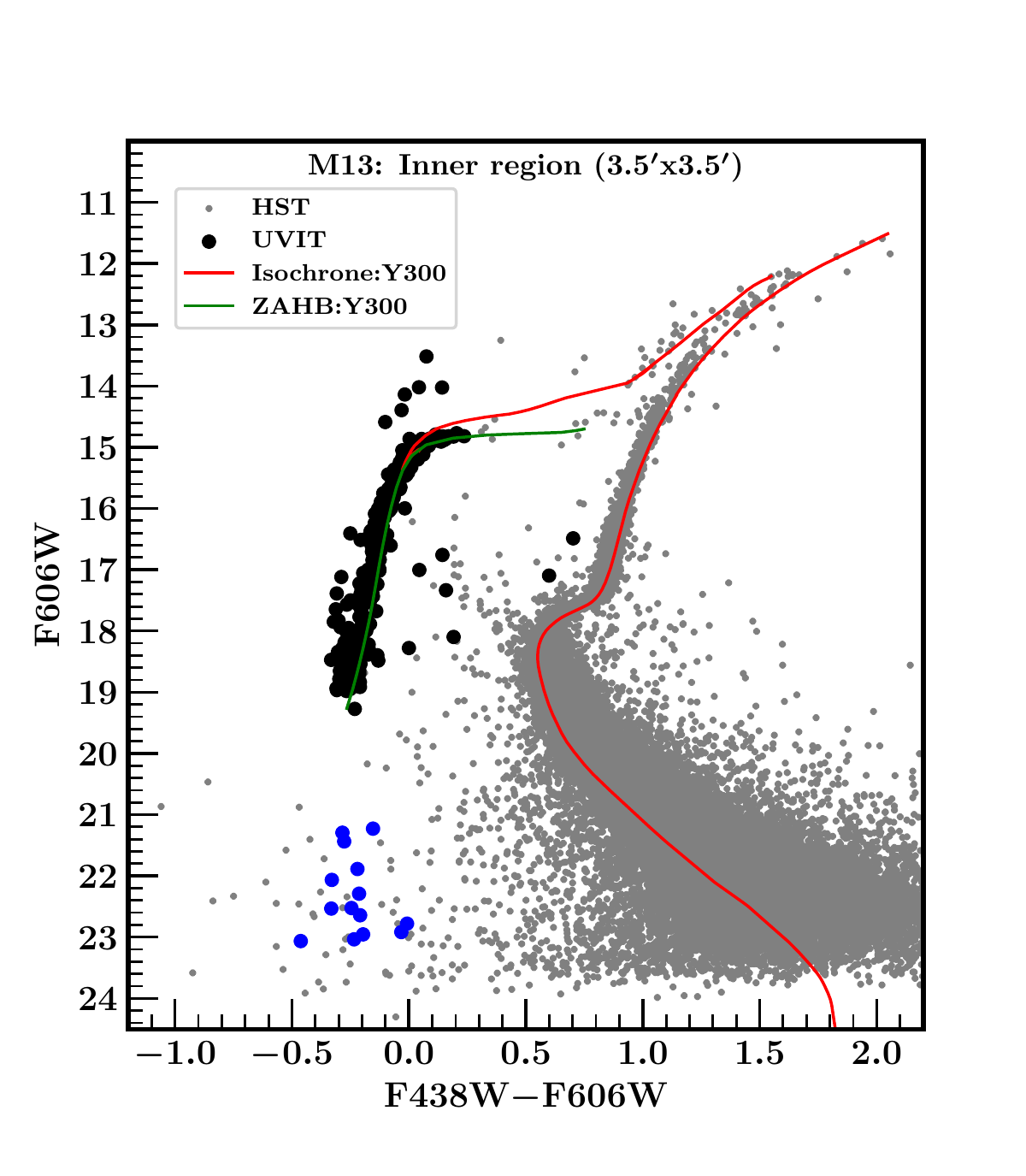}
    \includegraphics[width=0.49\textwidth]{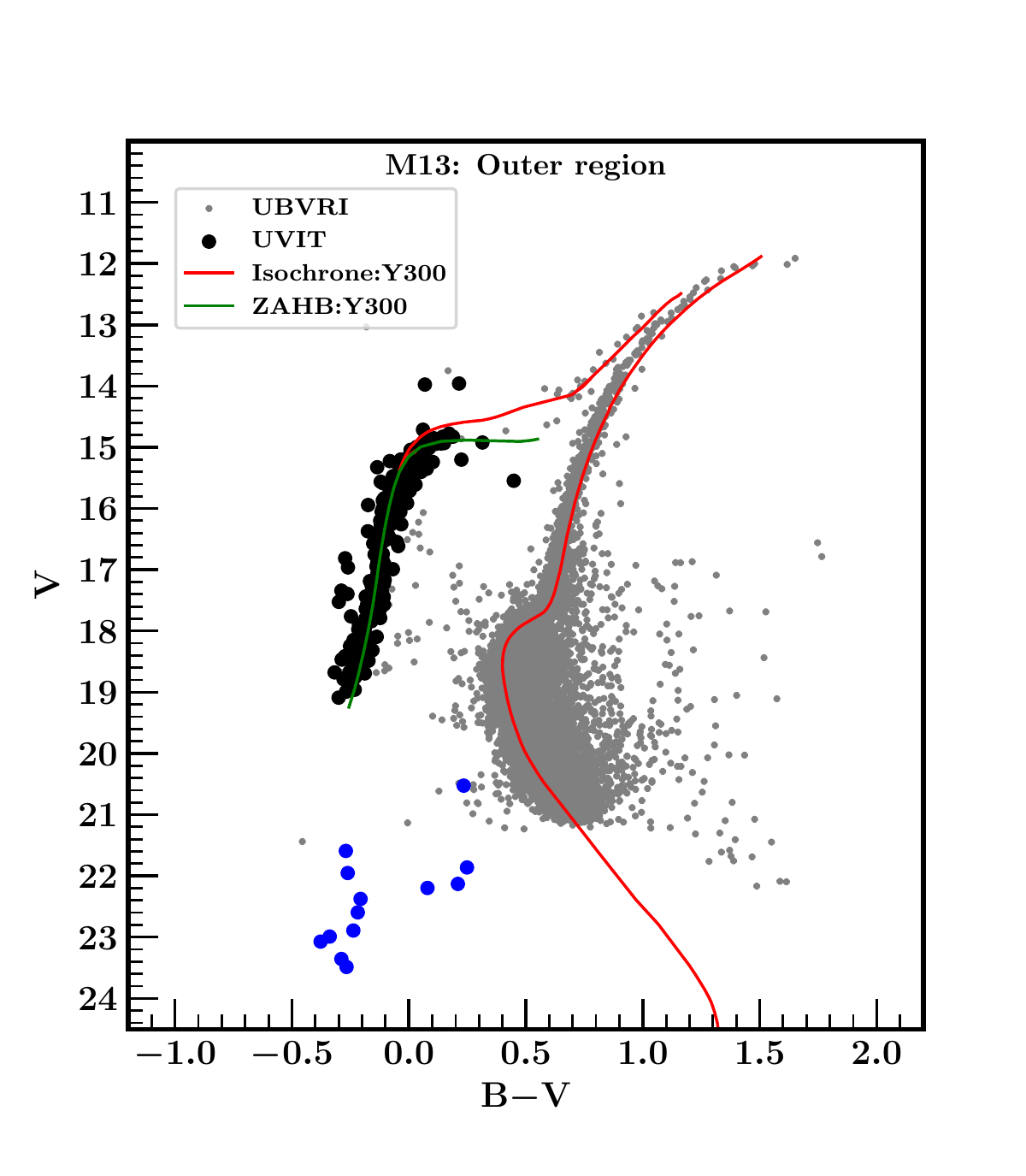}
\caption{Optical CMDs of M3 (upper panels) and M13 (lower panels) for \textit{HST} GCs catalogue (left panels) and UVBRI photometry data (right panels). The overall cluster members with membership probability $>90\%$ are shown in gray solids and the cross-matched UVIT counterparts are shown in black solids. UVIT observed WDs are denoted by blue solids. The red and blue coloured isochrones in the top panels are obtained for Y $=0.248$ and $0.275$, respectively, for GC M3 and the red coloured isochrones in the bottom panels are for Y $=0.300$ of M13. The ZAHB locus of ${\rm [Fe/H]}=-1.55$, ${\rm [\alpha/Fe]}=+0.4$, and Y $=0.300$ (green line) is shown in both the clusters. The physical parameters are taken from \citet{Denissenkov2017}.}
\label{fig:optical_cmd}
\end{figure*}

\begin{figure*}
    \centering
    \includegraphics[width=0.49\textwidth]{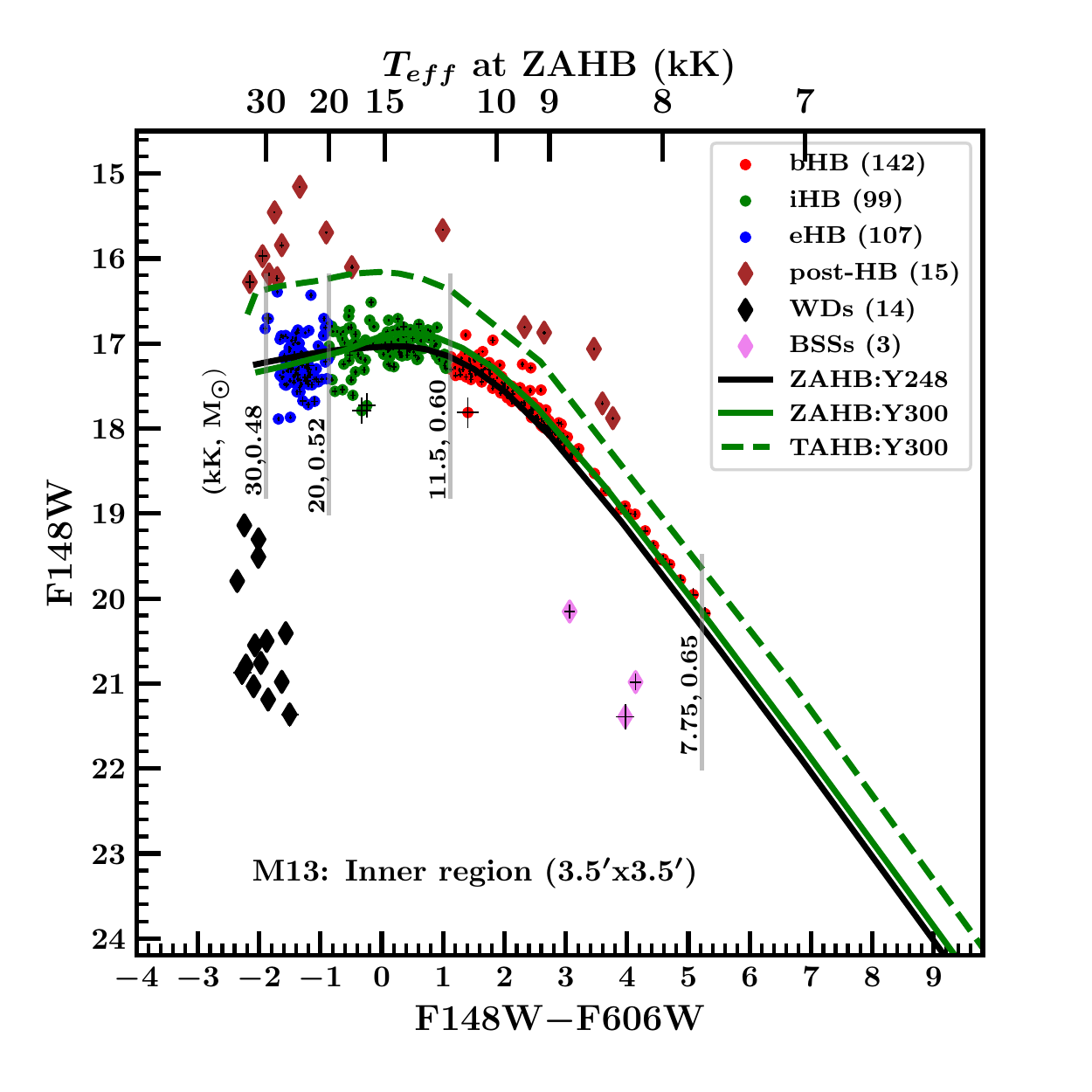}
    \includegraphics[width=0.49\textwidth]{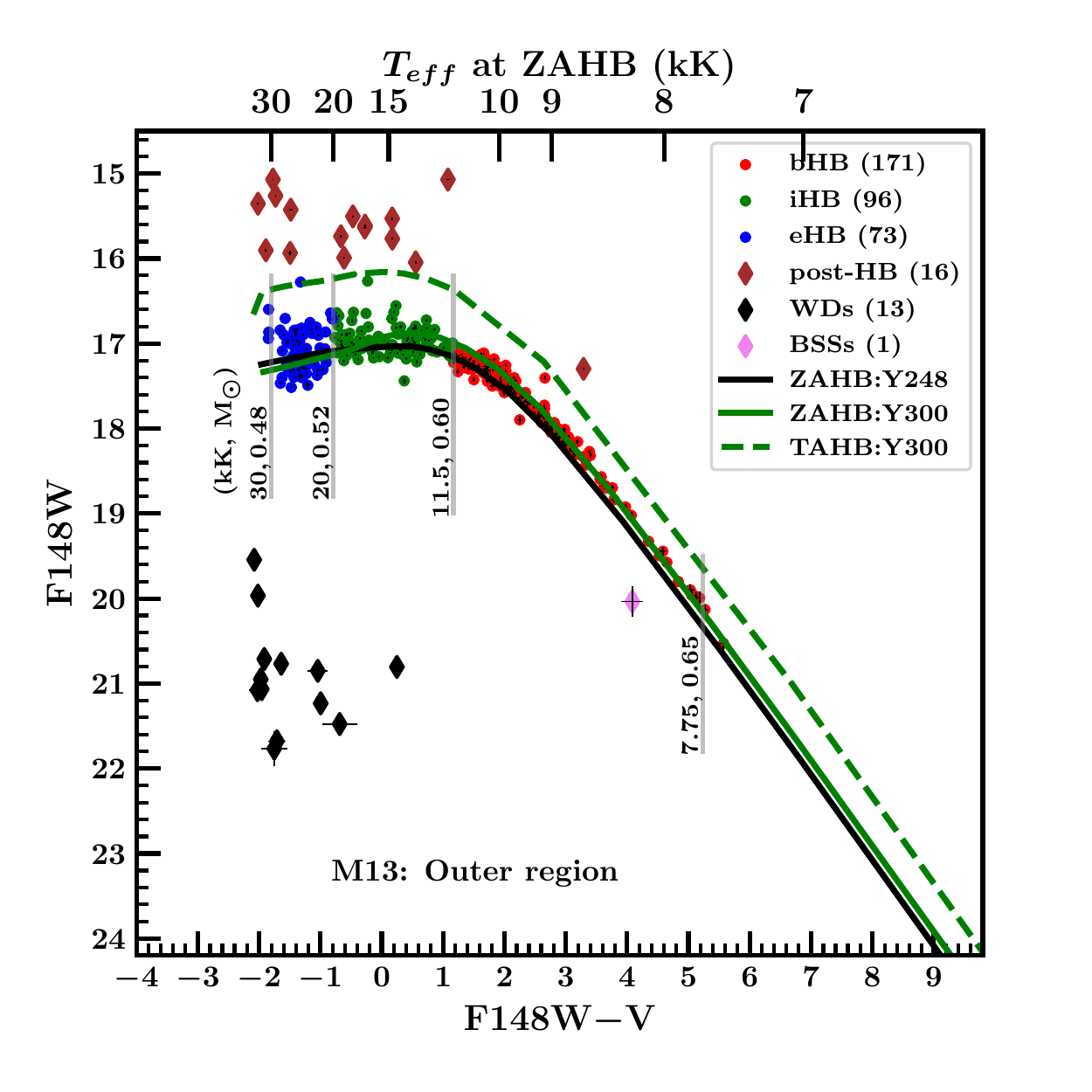}\\
    \includegraphics[width=0.49\textwidth]{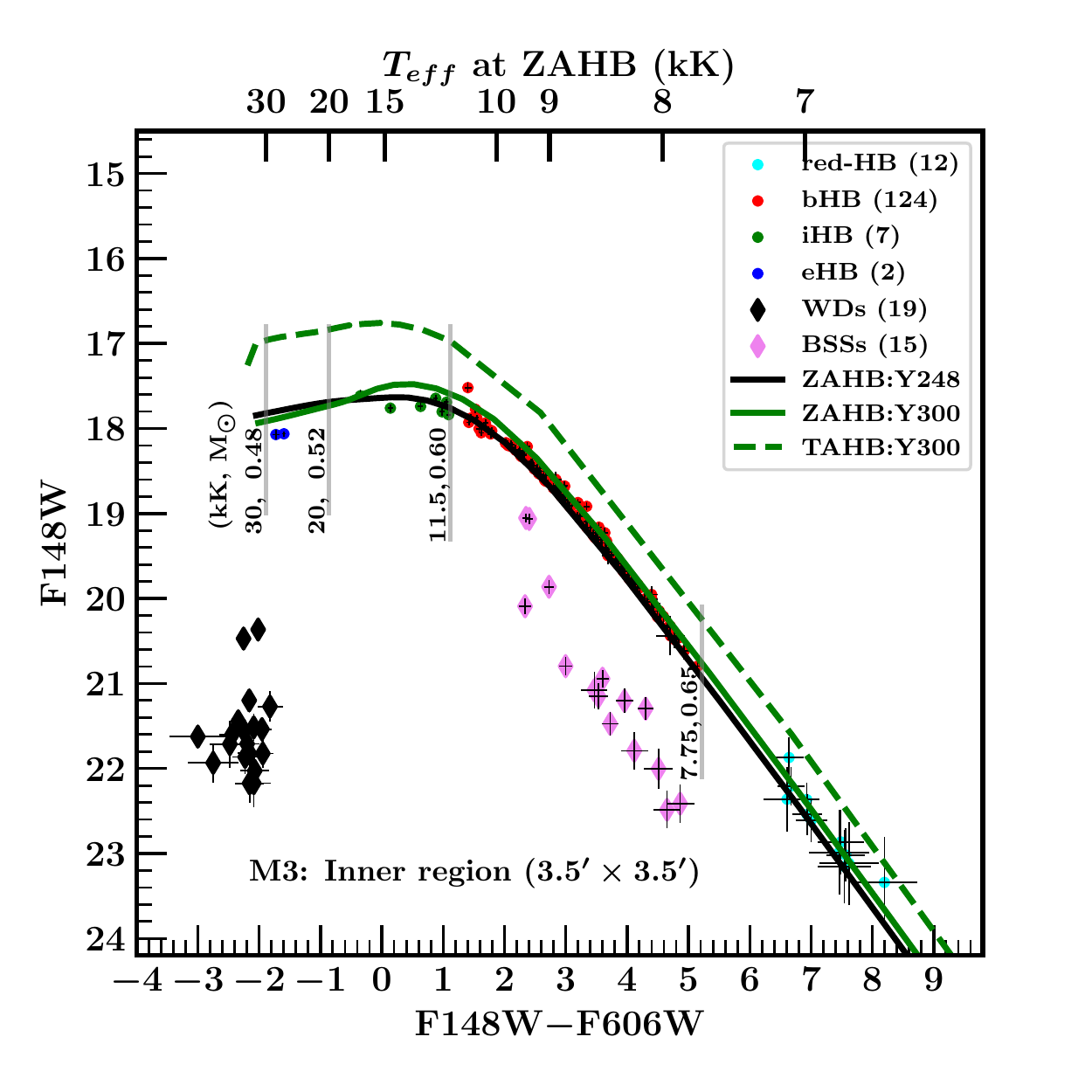}
    \includegraphics[width=0.49\textwidth]{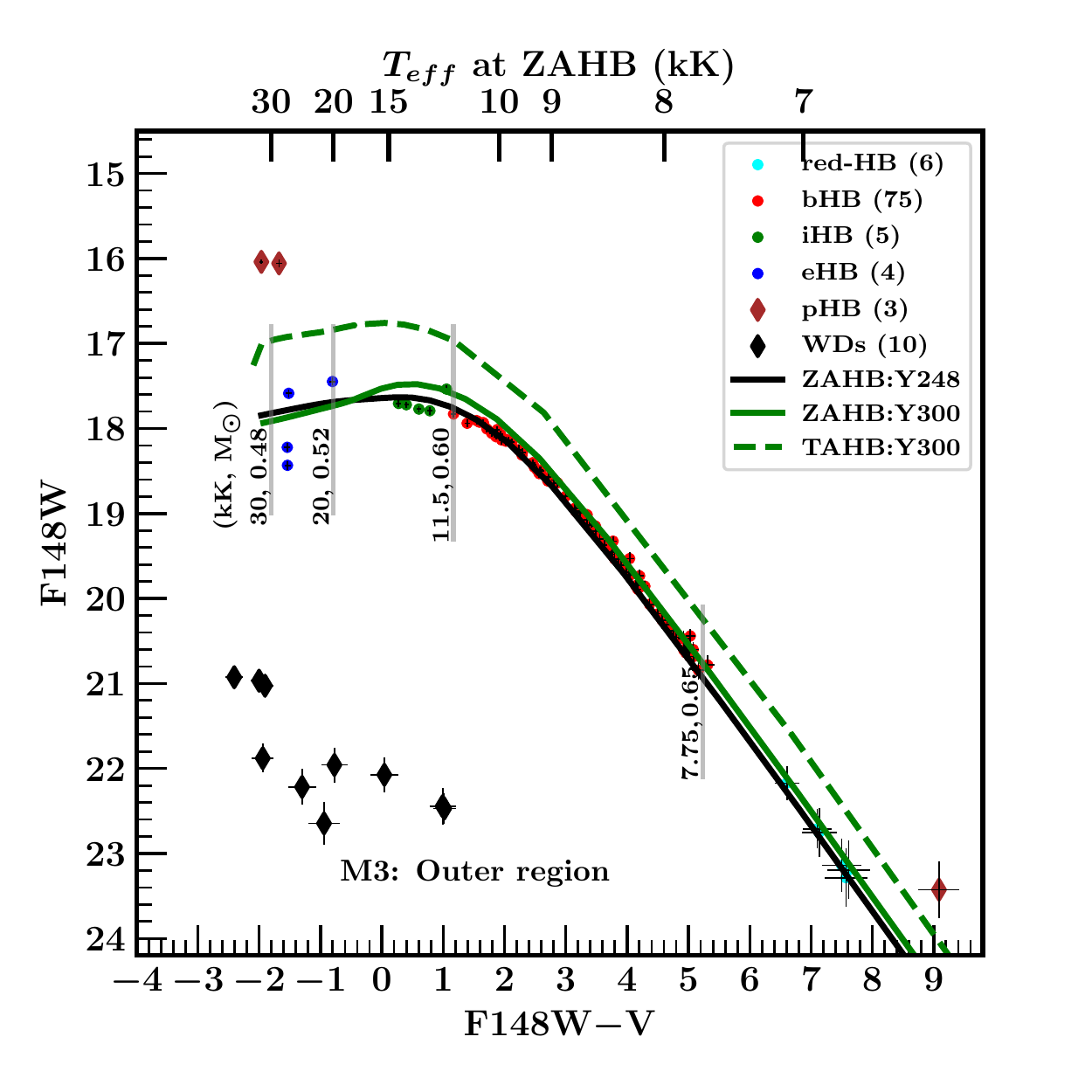}
    \caption{UV-optical CMDs of M13 (upper panels) and M3 (lower panels). The left panels show the CMDs of observed stars at the core (in the $3.5' \times 3.5'$ central region), and the right panels show the CMDs of the observed stars in the outer region of the clusters. Stars of different evolutionary phases (along with their respective counts) are shown by different symbols as mentioned in the legend. The ZAHB loci of Y $=0.248$ and Y $=0.300$ are represented by black and green solid lines. The TAHB locus of the Y $=0.300$ model is shown in the dashed green line. Different colour regions along the observed HB sequence are marked with vertical gray lines (see text for details). The \Teff\ (in kK) and \Mhb\ (in \Msun) of HB stars at the boundary of HB sub-populations are also indicated on the left of the vertical gray lines. The error bars in colour and magnitude of the observed stars are overlaid as black cross lines. }
    \label{fig:uv_cmd}
\end{figure*}

The optical CMDs for all the cluster members in the inner (within $3.5' \times 3.5'$ of the core and detected by {\em HST}) and outer (beyond $3.5' \times 3.5'$ of the core and observed by ground-based telescopes in UBVRI filters) regions of both the clusters are shown in \autoref{fig:optical_cmd} (M3: top panels and M13: bottom panels). The cluster members with membership probability $>90\%$ and observed with both {\em HST} (left panels) and UBVRI filters (right panels) are marked with gray solids. UVIT observed cross-matched sources are over-plotted in black solids. Sources shown in blue solids are FUV-detected sources for which the membership probability is not available (except a few, observed with {\em HST} in the inner region of the cluster), however, their positions in optical CMDs lie in the WD region, i.e., $3-4$ magnitudes fainter than main-sequence turn-off. Considering these sources as probable WD candidates, we discuss them later in this paper. The BaSTI-IAC isochrones and zero age HB (ZAHB) locus \citep{Pietrinferni2020} are over-plotted in the CMDs which are scaled to the observed plane using distance moduli of $15.02$ mag and $14.42$ mag for M3 and M13, respectively \citep{Denissenkov2017}.

We generated the isochrones and ZAHB locus for M3 using ${\rm [Fe/H]} = -1.55$ dex, ${\rm [\alpha/Fe]}=0.4$ dex, age $=12.6$ Gyr and two  different values of Y (0.248 and 0.275). As seen in \autoref{fig:optical_cmd}, the isochrone with Y = 0.248 (red solid line) is fitting well with the red HB region and the isochrone with Y = 0.275 (blue solid line) is lying in the blue HB region in CMDs of M3. But the isochrone with Y = 0.275 is not able to cover the magnitude spread of blue HB stars (i.e., the observed HB magnitude spread, $15.2-19.8$ ($14.8-19.2$) mag in the F606W filter for the inner region and $15.2-20.0$ ($14.8-19.2$) mag in the V filter for the outer region of the cluster M3 (M13)). However, the ZAHB locus with Y = 0.300 (green solid lines in the upper panels of \autoref{fig:optical_cmd}) is fitting well with the observed HB sequences in the inner as well as outer regions of M3. The bottom  panels of \autoref{fig:optical_cmd} show the optical CMDs for M13. The over-plotted isochrones and ZAHB locus in the figure are generated for ${\rm [Fe/H]}=-1.55$, ${\rm [\alpha/Fe]} = 0.4$, age $=12.9$ Gyr, and Y $=0.300$  which fit well with the observed optical CMDs.

\autoref{fig:uv_cmd} shows the UV-optical CMDs of GCs M13 and M3. We use UVIT F148W and {\em HST} F606W filters for the CMDs of inner regions whereas UVIT F148W and ground-based V filters for the CMDs of outer regions of the clusters. We superpose the BaSTI-IAC ZAHB loci for Y $= 0.248$ (black solid line), Y $= 0.300$ (green solid line) and terminal age HB (TAHB) locus for Y $= 0.300$ (green dashed line) on the CMDs. The ZAHB and TAHB loci are scaled to the observed plane as mentioned earlier. Labels in the figure represent the position of stars of different evolutionary phases. We notice a tight HB sequence along the ZAHB locus in both clusters. We assign stars above the TAHB locus (above the green dashed line) as post-HB stars (brown diamonds). The group of stars (black diamonds) located below the HB sequence are probable WDs. The stars below the HB sequence marked with violet diamonds are blue-straggler (BS) stars, as suggested in \citet{Schiavon2012} and \citet{sneha2020}. 

The observed HB sequence in \autoref{fig:uv_cmd} ranges from blue HB to extreme HB phase (${\rm 7,500\ K \leq T_{eff} \leq 30,000\ K}$) which has two prominent discontinuities (also known as gaps, jumps, over-luminous stars, or sub-luminous stars in different colours), the ``Grundahl jump" (G-jump) within the blue HB at $\sim11,500$ K \citep{Grundahl1998, Grundahl1999} and the ``Momany jump" (M-jump) within the extreme HB at $\sim20,000$ K \citep{Momany2002, Momany2004}. HB stars in between G-jump and M-jump (i.e., within \Teff\ range $11,500$ K to $20,000$ K) suffer from radiative levitation in the stellar atmosphere (increase in abundances of metal species like Fe, Cr, and Ti around solar or higher) and gravitational settling (depletion of He abundance by a factor of 10 or more) \citep{Behr2003, Moehler2003, Brown2016}. Hence, we divide the observed HB stars on the basis of temperature into three sub-populations -
\begin{center}
    blue-HB (bHB): ${\rm 7,750\ K \leq {\rm T_{eff}} \leq 11,500\ K}$ \\ \vspace{0.3 cm}
intermediate-HB (iHB): ${\rm 11,500\ K \leq {\rm T_{eff}} \leq 20,000\ K}$ \\ \vspace{0.3 cm}
extreme-HB (eHB): ${\rm 20,000\ K \leq {\rm T_{eff}} \leq 30,000\ K}$ \\
\vspace{0.3 cm}
\end{center}
We have used BaSTI-IAC ZAHB locus to co-relate the \Teff\ of HB stars with different UV-optical colours as shown in \autoref{fig:uv_cmd}. The F148W$-$F606W (F148W$-$V) colours at \Teff\ $7,750$ K, $11,500$ K, $20,000$ K, and $30,000$ K are 5.22 (5.23), 1.12 (1.17), $-0.86$ ($-0.79$), and $-1.89$ ($-1.81$), respectively and these colours or temperatures are marked in gray vertical lines on the HB sequence. We have also marked the corresponding HB masses (\Mhb) at these \Teff\ values in the CMDs. The \Mhb\ values corresponding to \Teff\ of $7,500$ K, $11,500$ K, $20,000$ K, and $30,000$ K are $0.65$ \Msun, $0.60$ \Msun, $0.52$ \Msun, and $0.48$ \Msun, respectively. The UVIT observed stars in bHB, iHB, and eHB regions are shown in red, green and blue solids, respectively. 

As observed in \autoref{fig:uv_cmd}, HB stars in M3 are mostly populated in the bHB region (with very few stars in iHB and eHB regions) whereas HB stars in M13 significantly populate the iHB and eHB regions (counts of HB stars in different sub-populations can be found in the legends of \autoref{fig:uv_cmd}). We also found a significant fraction of stars below the ZAHB locus in the eHB and iHB regions in M13. The observed fraction of stars below the ZAHB locus in the iHB region might be due to a higher Fe abundance than the cluster metallicity \citep{Moehler2003} whereas the fractional increase of eHB stars below ZAHB may be due to the fact that either they would be blue-hook (BHk) stars or eHB stars with higher Y than the considered Y for the plotted ZAHB in the figure \citep{Prabhu2021, sneha2020}. 

\section{Helium spread on HB stars of M3 and M13}    \label{sec:synth_hb}

\begin{figure*}
    \centering
    \includegraphics[width=0.49\textwidth]{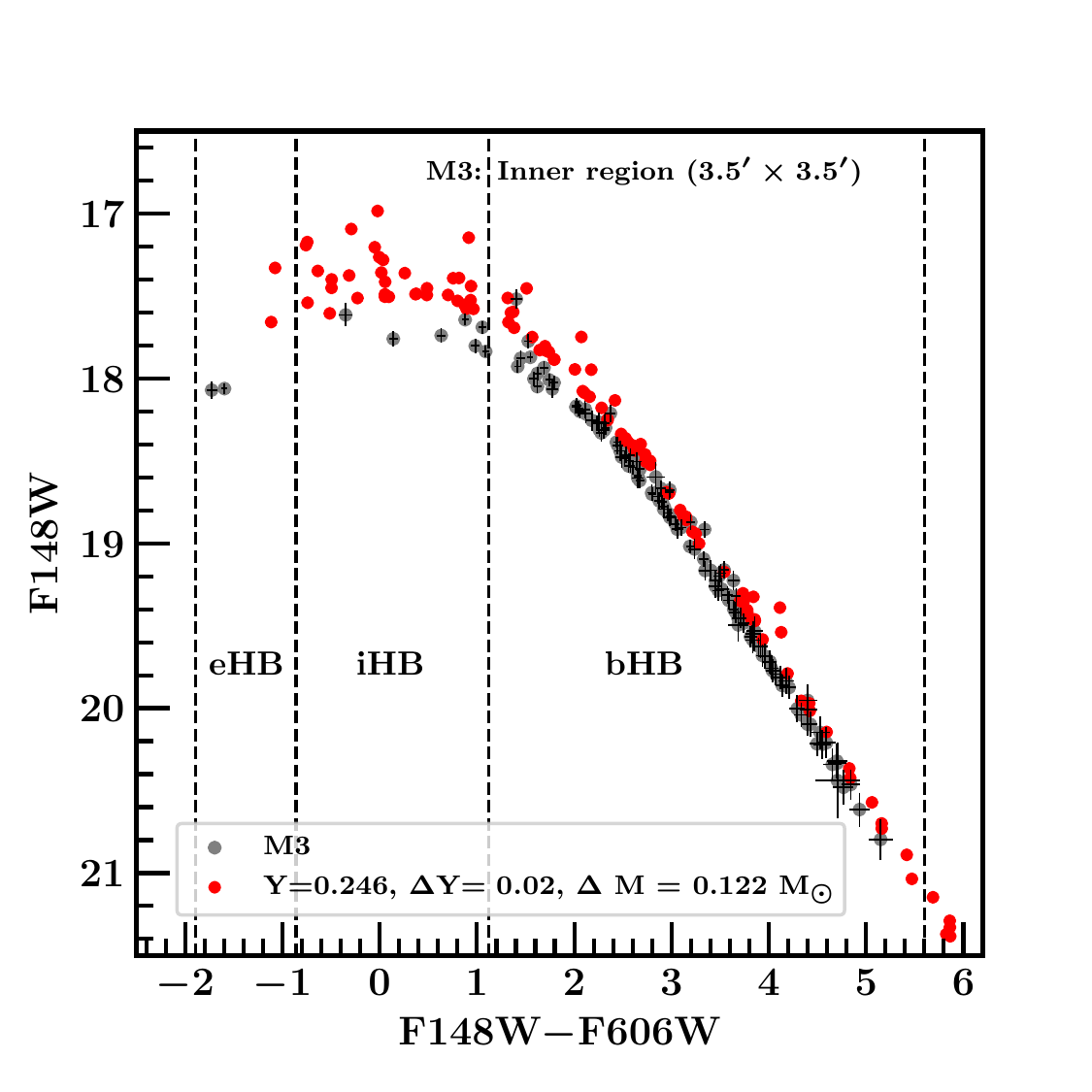}
    \includegraphics[width=0.49\textwidth]{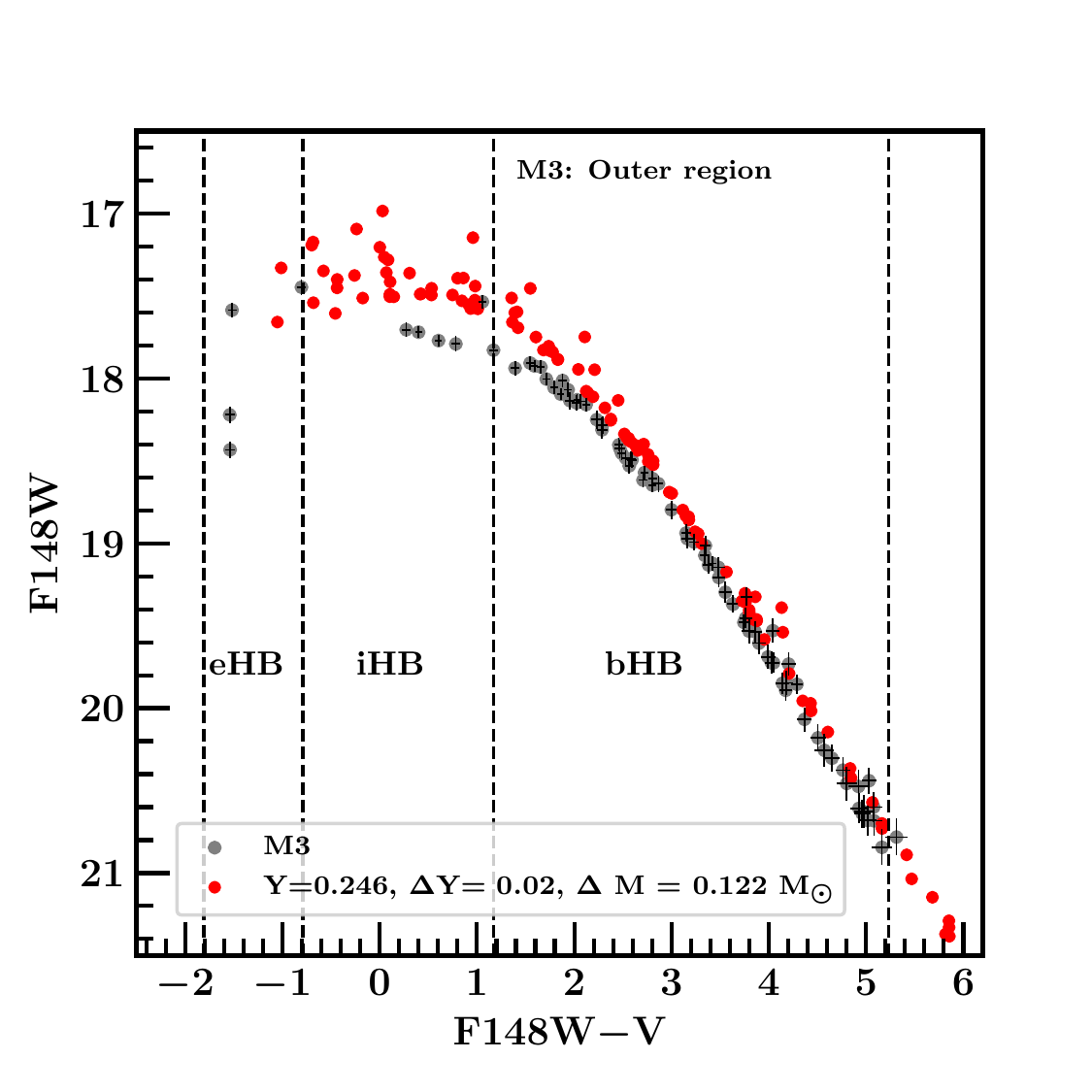}
     \includegraphics[width=0.49\textwidth]{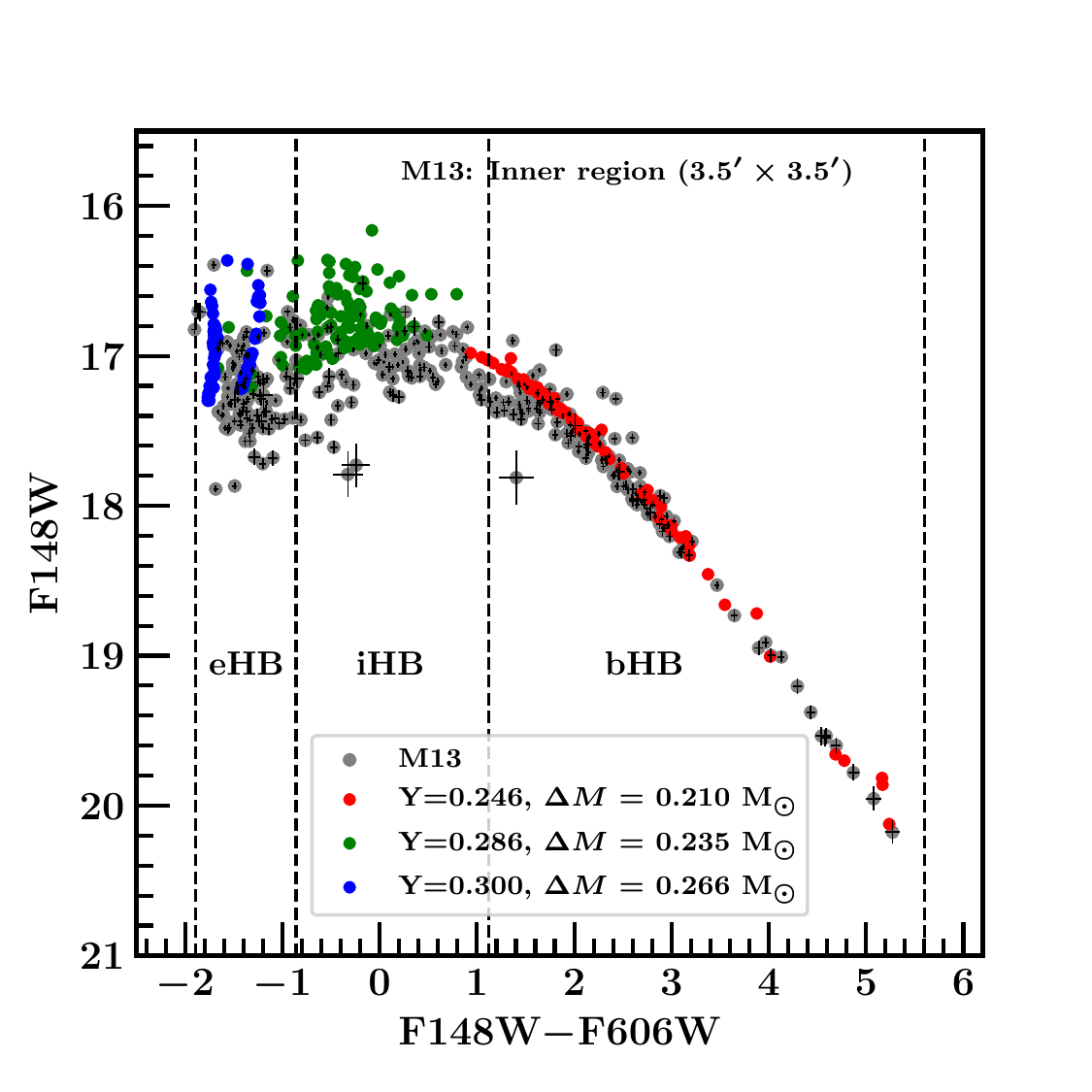}
    \includegraphics[width=0.49\textwidth]{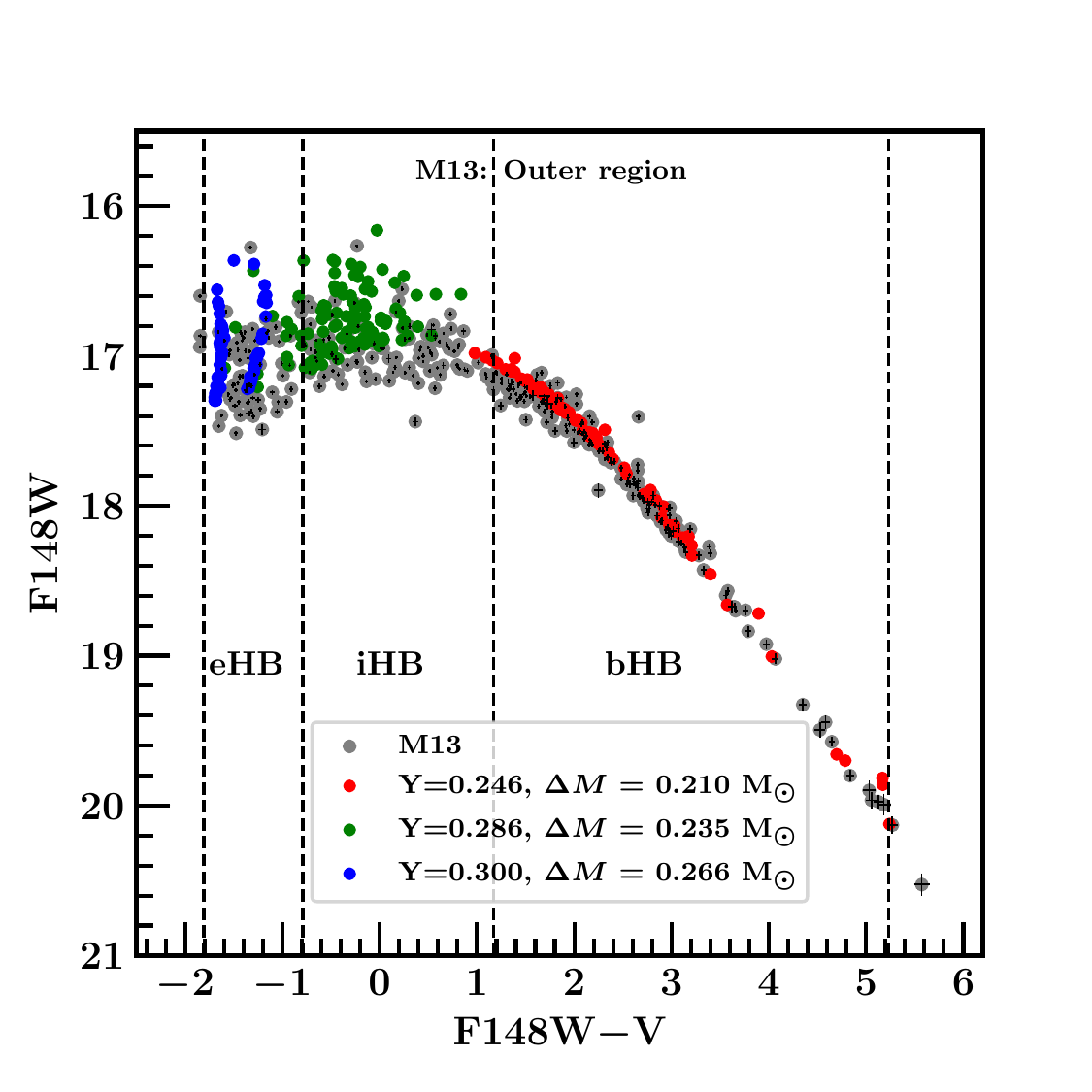}
    
    \caption{UV-optical CMDs of inner and outer regions of M3 and M13 as mentioned on the top of the respective panels. Gray solids represent the observed sources. Red solids in top panels are synthetic HB stars for Y$=0.246-0.265$ and mass-loss at RGB-tip ${\rm \Delta M} = 0.122\, {\rm M_\odot}$. Red, green, and blue solids in bottom panels are synthetic HB stars for Y and ${\rm \Delta M} =$ ($0.246$, $0.210\, {\rm M_\odot}$), ($0.286$, $0.235\, {\rm M_\odot}$), and ($0.300$, $0.266\, {\rm M_\odot}$), respectively. The  sub-populations of bHB, iHB, and eHB  are indicated by vertical dashed lines.}
    \label{fig:synth_m3_m13}
\end{figure*}

We have investigated the He-spread on the HB stars bluer than the RRL gap by comparing their synthetic CMDs with the observation as shown in \autoref{fig:synth_m3_m13}. We have adopted the methods provided by \citet{Dalessandro2013} to generate the synthetic magnitudes of HB stars in UVIT, {\em HST} and UBVRI filters. Calculation of synthetic magnitudes of HB stars requires specification of their chemical composition, He-spread, and mass-loss in the RGB phase.

The synthetic colours and magnitudes for HB stars in the M3 cluster are generated using ${\rm Y}=0.246$, $\Delta$Y $=0.02$ (uniform distribution), RGB total mass-loss $\Delta$M $ = 0.122$ \Msun\ with a spread in mass-loss governed by $\sigma$($\Delta$M) $=9.0$ × (Y$ - 0.246$) \Msun, where Y is the actual initial He mass fraction of the synthetic star. The dependence of $\sigma$($\Delta$M) on Y implies an increase of the mass-loss spread with increasing Y. In the upper panels of \autoref{fig:synth_m3_m13}, we have shown the synthetic and observed HB stars of M3 in red and gray solids, respectively. The observed colours of HB stars are matching well with the synthetic colours (F148W $-$ F606W for the inner and F148W $-$ V for the outer regions of the cluster) in  the bHB region, whereas there is a deviation between them in eHB and iHB regions. 

Similarly, the synthetic colours and magnitudes for HB stars in M13 are generated using the best-fitted Y and $\Delta$M parameters derived by \citet{Dalessandro2013} which covers the following ranges of Y and ${\rm \Delta M}$:\\ 
\noindent (i) Y $= 0.246$ to $0.256$ (uniform distribution), and $\Delta$M $=0.21$ \Msun\ with $\sigma$($\Delta$M) $=0.01$ \Msun\ (Gaussian distribution): denoted by red solids in the lower panels of \autoref{fig:synth_m3_m13}.\\
\noindent (ii) Y $= 0.285 \pm 0.012$ (Gaussian distribution), and $\Delta$M $= 0.235$ \Msun\ with $\sigma$($\Delta$M) $= 0.01$ \Msun\ (Gaussian distribution): denoted by green solids in the lower panels of \autoref{fig:synth_m3_m13}.\\ 
\noindent (iii) Y $= 0.300 \pm 0.003$ (Gaussian distribution), and $\Delta$M $= 0.266$ \Msun\ with $\sigma$($\Delta$M) $= 0.002$ \Msun\ (Gaussian distribution): denoted by blue solids in the lower panels of \autoref{fig:synth_m3_m13}. 
We noticed that the observed colours (gray solids) match well with the synthetic colours in the bHB region whereas there is a deviation between the colours in both iHB and eHB regions. Almost $40\%$ of the observed eHB and iHB stars are slightly fainter in F148W magnitudes than the corresponding synthetic values. 

The observed and synthetic F148W $-$ F606W colours in the lower panels of \autoref{fig:synth_m3_m13} suggest that bHB stars in M13 are distributed with Y in the range $0.246$ to $0.256$. The iHB stars are distributed with Y in the range $0.270$ to $0.300$ (green solids). The eHB stars have He-distribution in the range of $0.290$ to $0.310$ (blue solids). \citet{Denissenkov2017} have also studied the HB stars of M3 and M13 GCs using synthetic colours of optical filters. However, they used synthetic HB stars with a relatively larger spread in Y than in ${\rm \Delta M}$. Their best fit models for M3 suggest ${\rm Y = 0.250,\ \Delta Y \sim 0.01,\ and\ \Delta M = 0.13\ M_\odot}$. Whereas synthetic HB models used in this paper for M3 are ${\rm Y = 0.246,\ \Delta Y \sim 0.02,\ and\ \Delta M = 0.122\ M_\odot}$. These simulations are able to produce similar colours and magnitudes in FUV-optical CMDs as observed in bHB stars of M3. The best-fit models in \citet{Denissenkov2017} for M13 suggest ${\rm Y = 0.250,\ \Delta Y \sim 0.08,\ and\ \Delta M = 0.2\ M_\odot}$. Whereas synthetic HB models used in this paper for M13 are ${\rm Y = 0.246,\ \Delta Y \sim 0.06,\ and\ \Delta M = 0.266\ M_\odot}$. The observed HB stars in M13 in UV-optical CMDs cover the colour range, however, the UV magnitudes of simulated HB stars are brighter than the observed HB stars in eHB and iHB regions. The comparison of synthetic and observed colours of HB stars suggests that the mass-loss at RGB and initial He-spread in both clusters have a simultaneous effect on the different HB distributions observed in M3 and M13. e.g., HB stars of M13 require a larger spread in Y as well as ${\rm \Delta M}$. M13 with a larger fraction of eHB and iHB stars than M3 can only be simulated if HB distribution has higher values of Y and ${\rm \Delta M}$.

\subsection{Multiple populations along RGB and their impact on HB morphology of M13 and M3}

The chromosome maps of M13  have revealed that it consists of a long sequence of 2G stars and a small number of 1G stars along the RGB \citep[Figure 5 of ][]{Milone2017} with a fraction of $0.184 \pm 0.013$ 1G RGB stars. The UVIT observed HB sequence of M13 starts from the bluer side of the RRL stars (with a small number of RRL stars residing towards the bluer end of the instability strip; see Figure \ref{fig:optical_cmd}). This suggests that $\sim$18\% of the observed blue HB stars (including RRL stars) would be the progeny of the 1G RGB stars. Hence, to reproduce them, initial He-abundance Y$\sim$0.25 dex  and respective RGB mass-loss ${\rm \Delta M = 0.210\ M_\odot}$ of 1G stars would be required \citep{Milone2018, Tailo2020}. A similar initial He abundance and mass-loss values have been used in \cite{Dalessandro2013} and also in our present work, i.e, Y$\sim$ 0.246 to 0.256 dex and ${\rm \Delta M = 0.210\ M_\odot}$ to reproduce the blue HB stars near the RRL strip.

The eHB stars (the bluest and the hottest HB stars) were reproduced by \cite{Tailo2020} using an enhanced He-abundance (${\rm \delta Y_{max} = 0.052\pm0.004}$) and a higher mass-loss at RGB (${\rm \Delta M = 0.273\pm0.021\ M_\odot}$). \cite{Milone2018} suggested that such an extreme He enrichment represents the extreme 2G (2Ge, bluest 2G RGB) stars in the chromosome map \citep[Figure 2 of ][]{Milone2018}. Hence, the extreme HB stars would be the progeny of 2Ge RGB stars. \cite{Dalessandro2013} have used a similar prescription of He enhancement (${\rm \Delta Y = 0.05}$) and mass-loss along RGB stars ${\rm \Delta M = 0.266\ M_\odot}$) to reproduce the extreme HB stars of M13 and we have used their values in our work.

A comparison between the observed and simulated HB sequences (lower panels of \autoref{fig:synth_m3_m13}) suggests that the overall observed UV-optical colours are covered by the simulated stars with three different He and RGB mass-loss prescriptions. However, the observed and simulated HB number counts are not matching well when using the Gaussian distribution of He-abundance (with ${\rm Y_{mean} = 0.285}$ and $0.300$) and RGB mass-loss (${\rm \Delta M_{mean} = 0.235\ M_\odot}$ and $0.266$ \Msun). Hence, we suggest a uniform distribution of He and mass-loss along RGB would be more efficient in producing the observed star counts along the HB sequence.

\begin{figure}
    \centering
    \includegraphics[width=\columnwidth]{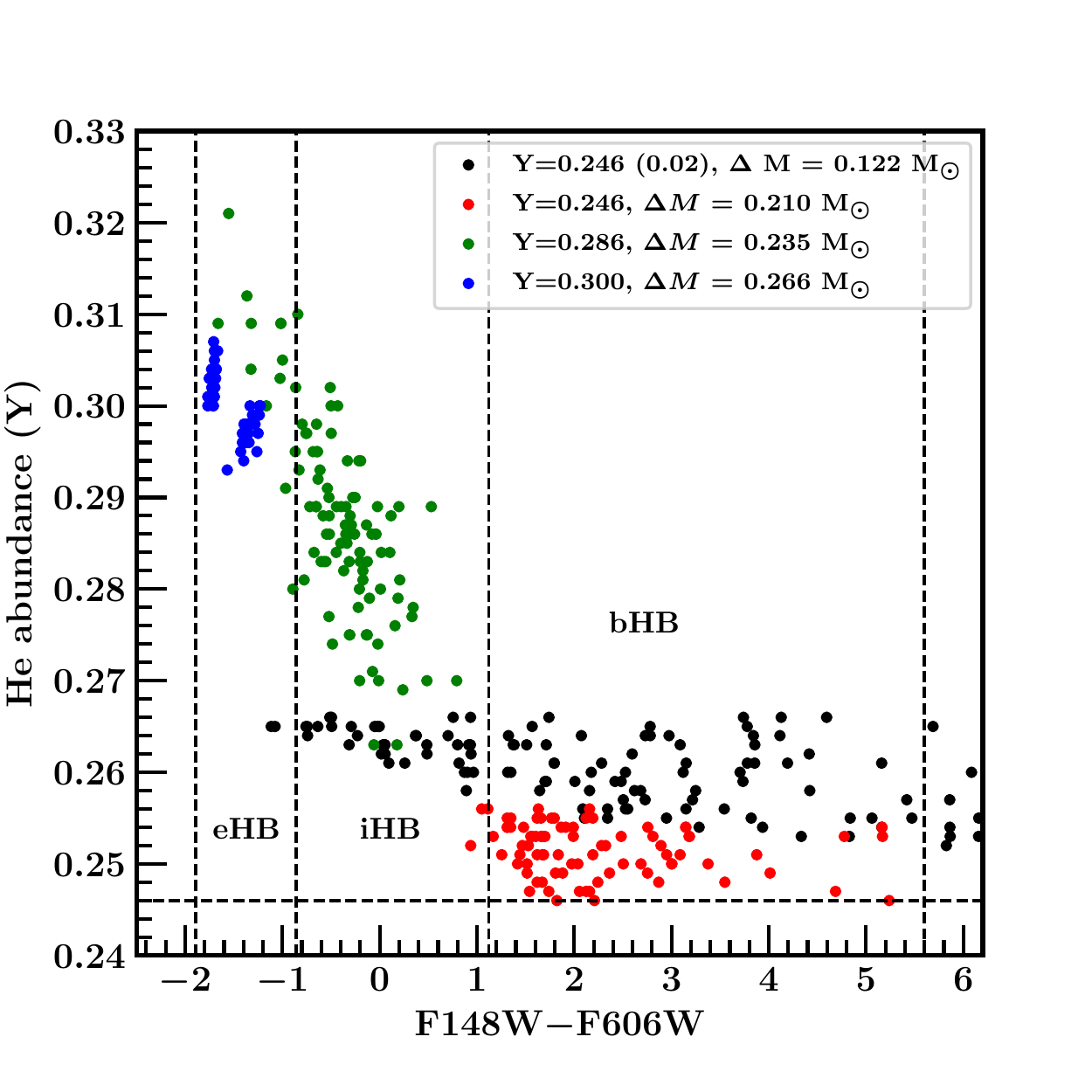}
    \caption { F148W $-$ F606W vs. He-spread plot for synthetic HB stars. Synthetic HB stars used to compare the observed HB stars in M13 are shown in red, green, and blue solids as described in the legends for different combinations of He-spread and mass-loss on RGB. The black solids are synthetic HB stars used to compare the observed HB stars of M3. bHB, iHB, and eHB sub-populations are indicated by vertical dashed lines. The horizontal dashed line shows the canonical He-abundance (Y$_{init}$ = 0.247). }
    \label{fig:synth}
\end{figure}

The average and maximum value of He enhancement between 1G and 2G RGB stars of M3 as estimated by \cite{Milone2018} are ${\rm \Delta Y_{1G,2G}}$ = 0.016$\pm$0.005 and ${\rm \delta Y_{max}} =0.041\pm0.009$, respectively. These values are used by  \cite{Tailo2019} to estimate the mass-loss of RGB stars to reproduce the HB stars of M3. They divided the observed blue HB stars of M3 into three groups: ${\rm 2G_A}$, ${\rm 2G_B}$, and ${\rm 2G_C}$ with the ${\rm 2G_A}$ group having the greatest amount of blue HB stars. The assigned He enhancement for ${\rm 2G_A}$ stars are ${\rm \Delta Y_{1G,2G}}$ (${\rm Y_{1G} = 0.25 }$), ${\rm 2G_C}$ stars are ${\rm \delta Y_{max}}$, and ${\rm 2G_B}$ stars are in the range between ${\rm \Delta Y_{1G,2G}}$ and ${\rm \delta Y_{max}}$. Their best fitted results indicate RGB mass-loss of 0.204 \Msun, 0.220 \Msun, 0.240 \Msun\ in ${\rm 2G_A}$, ${\rm 2G_B}$, and ${\rm 2G_C}$ HB stars, respectively. 
The blue HB stars of M3 that we study here are split into three categories (bHB, iHB, and eHB stars). \autoref{fig:synth_m3_m13} and \ref{fig:synth} show that the He-spread of 0.252 to 0.266 (${\rm \Delta Y}$ = 0.16 taking ${\rm Y_{1G}}$ = 0.25) is a good fit for the bHB stars and is similar to the ${\rm 2G_A}$ population fitted by \cite{Tailo2019}. The iHB stars of M3 match well with a He-abundance of 0.258 to 0.266, which is comparable to the ${\rm 2G_B}$ HB population identified by \cite{Tailo2019}, and can be used to fit the iHB stars of M3. To replicate the eHB stars of M3 that mirror the ${\rm 2G_C}$ population in \cite{Tailo2019}, it is unquestionably necessary to increase He-abundance and mass-loss values. Here, we note that we have used relatively lower values of mass-loss (${\rm \Delta M = 0.122}$ \Msun) and a spread in mass-loss using equation ${\rm \sigma(\Delta M) = 9.0\times (Y-0.246)}$ in the blue HB stars as compared to \cite{Tailo2019}. The simulated HB stars match well with the observed colour spread in the bHB stars of M3 (\autoref{fig:synth_m3_m13}).
 
Recently, it has been shown that in many GCs there is a spread in the [Fe/H] abundances in the 1G RGB stars \citep{Marino2019, Lardo2022, Legnardi2022, Lardo2023} as obtained on the basis of the colour distribution of stars in the chromosome map. However, the UVIT observed HB stars of M3 and M13 are blue HB stars which are the progeny of the 2G RGB population, hence the variation in their metallicity should be small as stated by \cite{Legnardi2022}. Hence, the iron spread in the 1G RGB population of the clusters will not affect the present analysis of blue HB stars of M3 and M13.

\section{Hot post-HB stars in M3 and M13} \label{sec:post_hb}

The post-HB stars when evolving from eHB and iHB regions become brighter in UV filters and appear as a vertical sequence above eHB and iHB stars in the UV-optical CMDs \citep{Schiavon2012, Dalessandro2013, Moehler2019, Prabhu2021}. Although  \citet{Dalessandro2013} indicated the presence of post-HB stars in M13 using {\em HST} data, they did not provide any detailed statistics about them. We have used F148W$-$F606W vs F148W CMD for the inner region (left panels of \autoref{fig:uv_cmd}) and F148W$-$V vs F148W CMD for the outer region (right panels of \autoref{fig:uv_cmd}) of the clusters to identify post-HB stars. We identified a total of 31 post-HB stars in M13 and 3 post-HB stars in M3 which are brighter than the observed HB stars in F148W filters at the similar (F148W$-$F606W)/(F148W$-$V) colours. There are 15 (16) post-HB stars in the inner (outer) region of M13. They all appear to be distributed above the eHB, iHB, and bHB stars. Recently, \citet{Davis2021} have reported 13 post-HB stars in M13 considering stars above HB region, spanning from red HB to bHB, in optical (V$-$I vs. V) CMDs. Such stars have originated from bHB or red HB stars of the cluster \citep{Moehler2019}. However, we see a possibility of a large number of post-HB stars evolving from the populated eHB and iHB in M13. \citet{Chen2021} have mentioned only about 7 post-HB stars evolving from the eHB region while studying the inner region of the cluster with the {\em HST} F275W filter (which is a near-UV filter). We have detected all the 7 reported post-HB stars in \citet{Chen2021} along with additional 24 post-HB stars in the inner as well as outer regions of M13.

In M3, we could observe post-HB stars only in the outer region of the cluster. Out of the 3 post-HB stars in M3, 2 are above the eHB, and 1 above the red HB. This suggests that only 2 post-HB stars are evolving from the eHB in M3. \citet{Schiavon2012} have found 2 AGB-manqu\'e and 1 PEAGB stars in M3. The 2 post-HB stars identified above eHB are the 2 AGB-manqu\'e stars detected by \citet{Schiavon2012}. The PEAGB star of \citet{Schiavon2012} is saturated in UVIT images hence we had to exclude it from our analysis.

\subsection{Evolutionary status of post-HB stars in M3 and M13} \label{sec:evol_status}

\begin{figure*}
    \centering
    \includegraphics[width=0.49\textwidth]{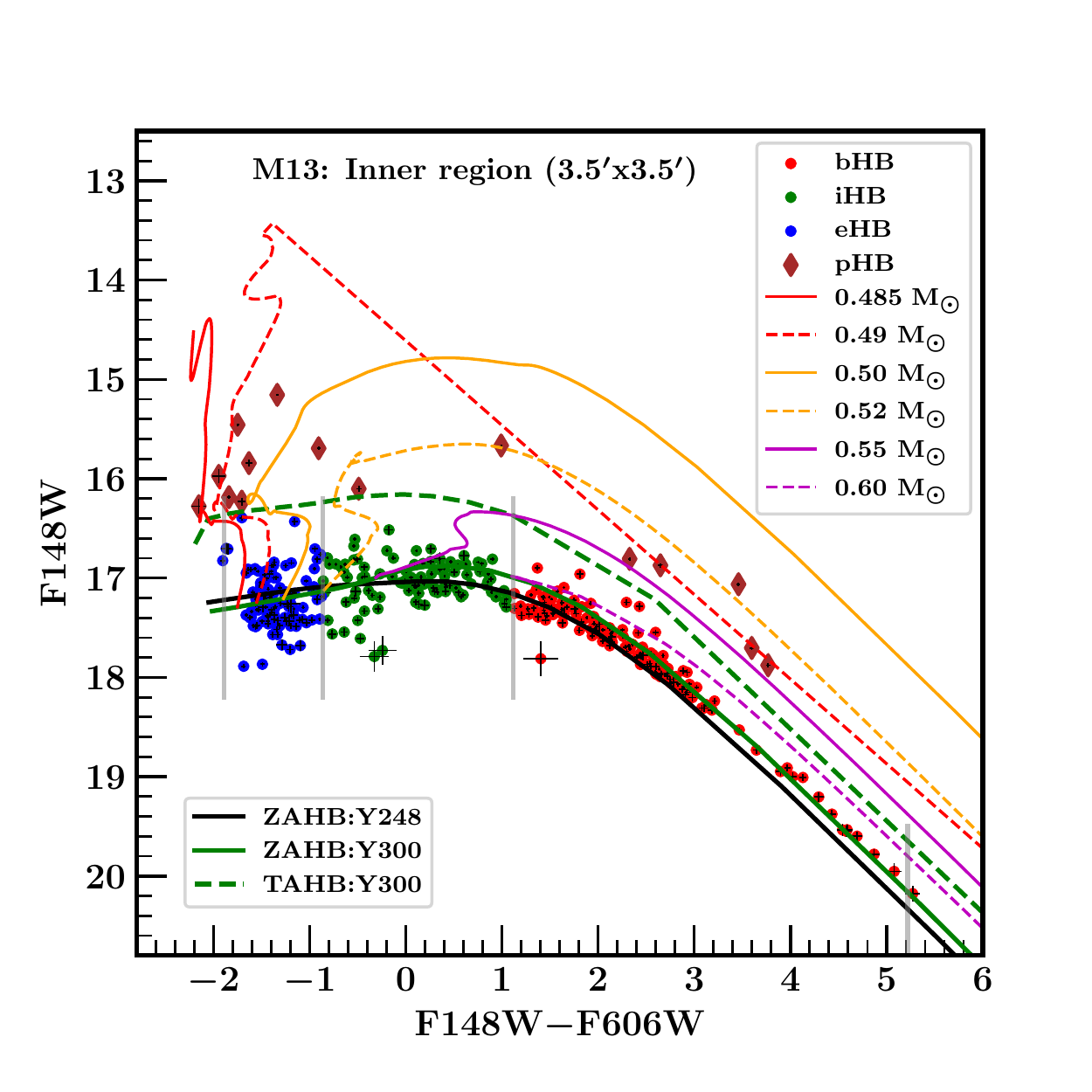}
    \includegraphics[width=0.49\textwidth]{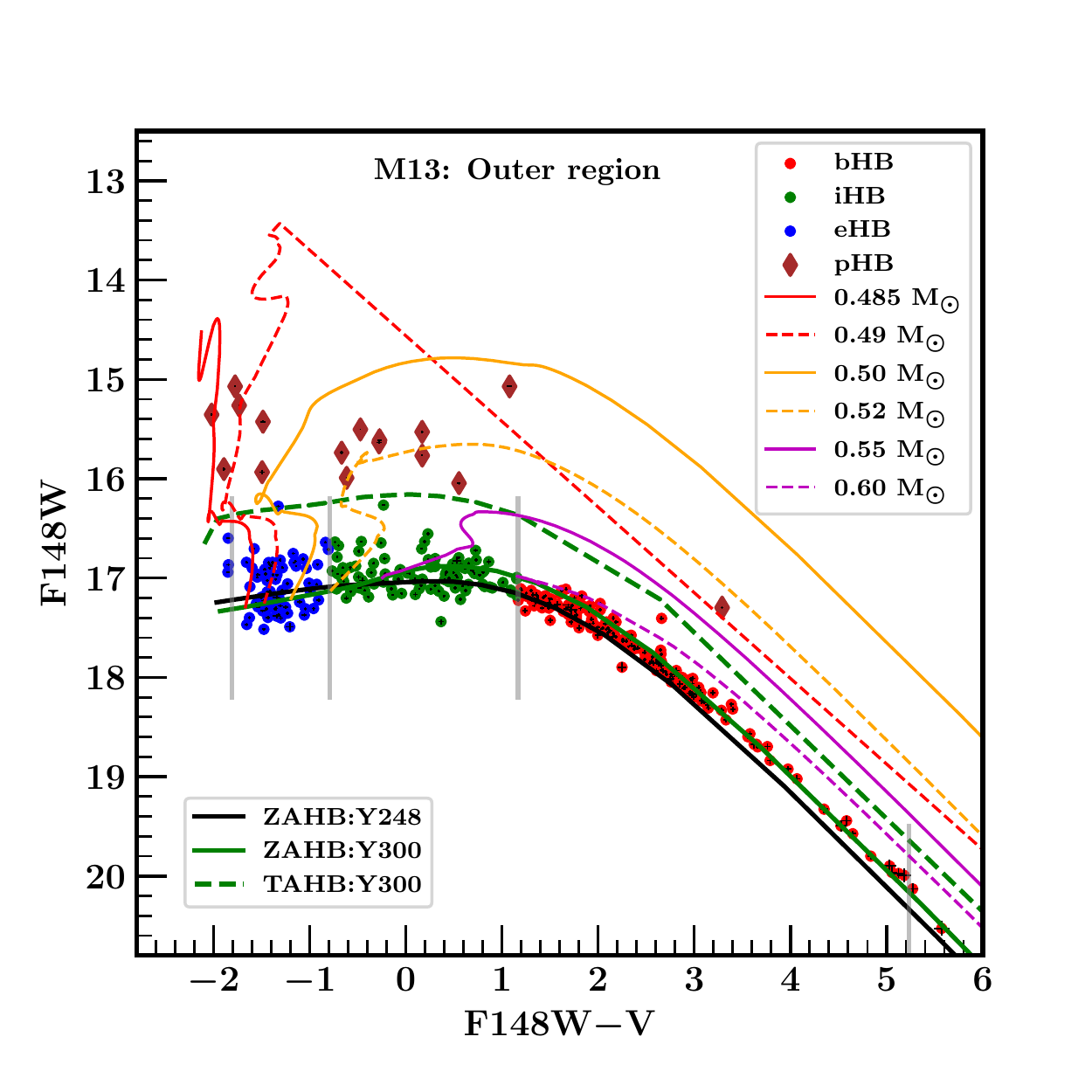}
    \includegraphics[width=0.49\textwidth]{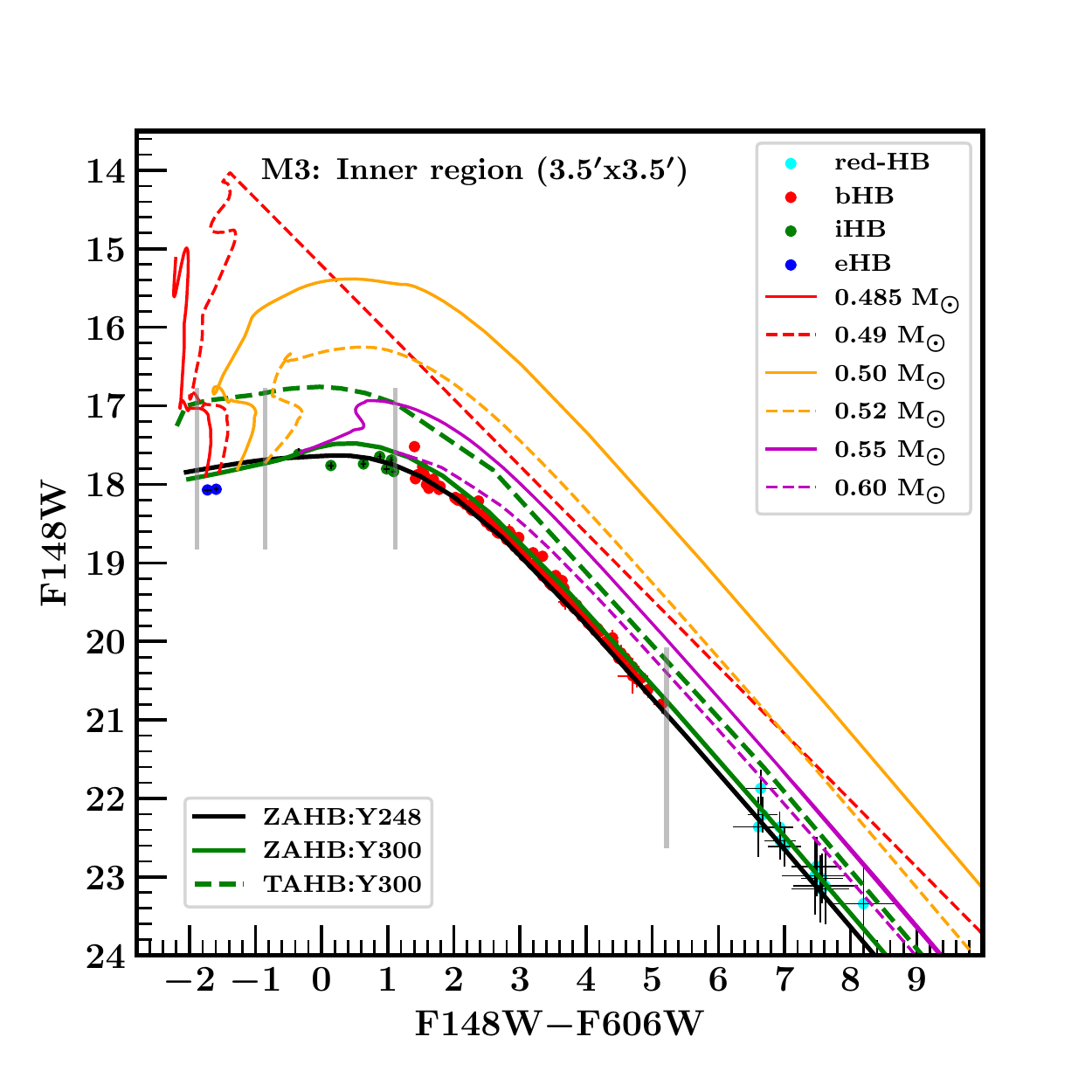}
    \includegraphics[width=0.49\textwidth]{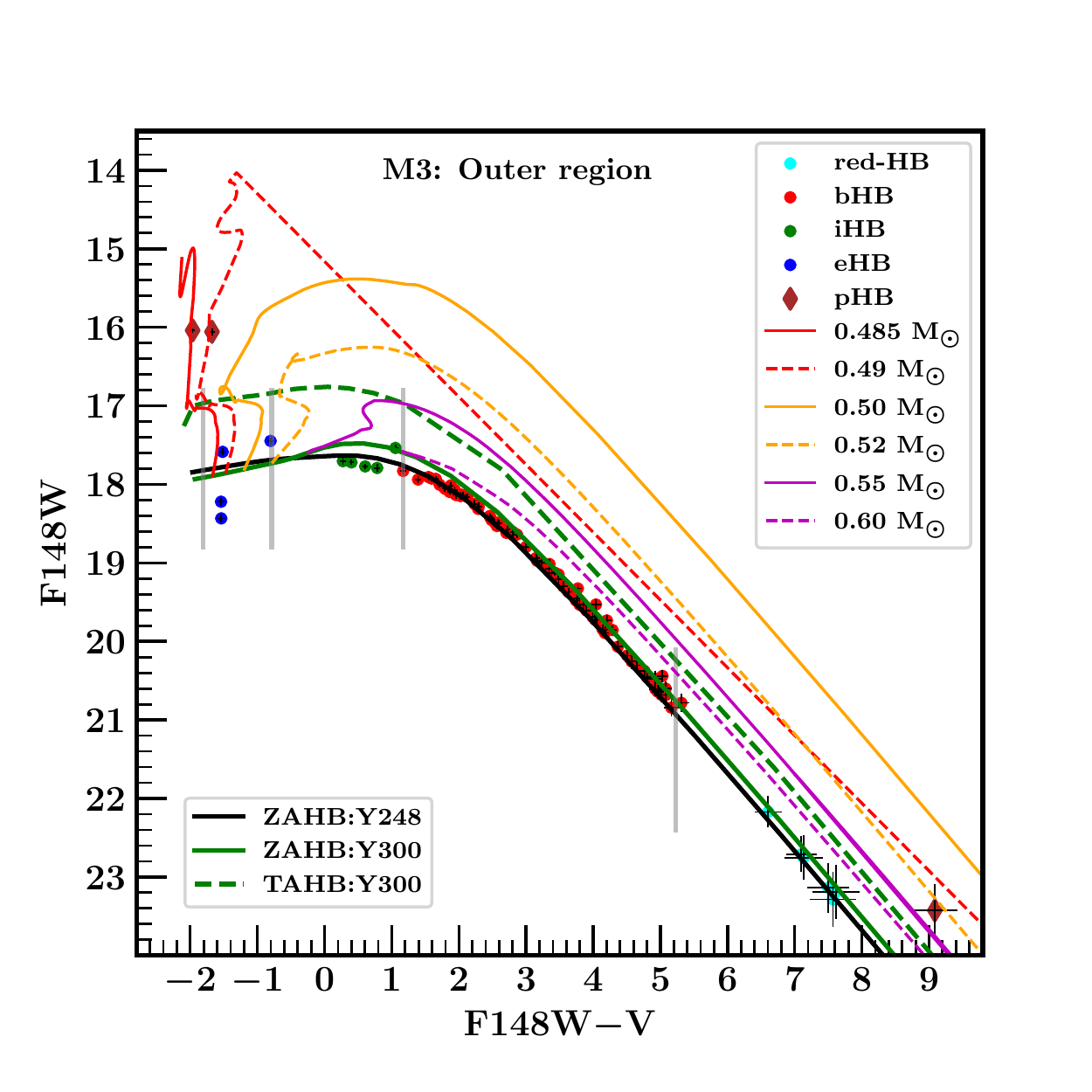}
    \caption{UV-optical CMDs of the UVIT observed sources of inner and outer regions of M13 and M3 as mentioned on the top of the respective panels. Post-ZAHB evolutionary tracks from the BaSTI-IAC model are for ZAHB mass of 0.485 \Msun\ (red solid line), 0.49 \Msun\ (red dashed line), 0.50 \Msun\ (yellow solid line), 0.52 \Msun\ (yellow dashed line), 0.55 \Msun\ (violet solid line), and 0.60 \Msun\ (violet dashed line). The green and black solid lines are ZAHB locus of initial He-abundance ${\rm Y=0.248 \ and\ Y=0.300}$ respectively. The  green dashed line represents the locus of TAHB.}
    \label{fig:cmd_evol}
\end{figure*}

\begin{table*}
    \centering
    \caption{Number of post-HB stars in M3 and M13 evolved from ZAHB locus with ${\rm M_{HB}} \leq 0.55{\rm M_\odot}$.}
    \label{tab:pHB}
    \resizebox{\textwidth}{!}{
    \begin{tabular}{lccccc} \hline
          \multirow{2}{*}{Cluster name} & \multirow{2}{*}{region} & \multicolumn{4}{c}{ Number of post-HB stars evolved from different HB mass range}  \\
        & & ${\rm M_{HB} \leq 0.49\ M_\odot}$ & $ {\rm 0.49\ M_\odot \leq M_{HB} \leq 0.50\ M_\odot}$ & ${\rm 0.50\ M_\odot \leq M_{HB} \leq 0.52\ M_\odot}$ & ${\rm 0.52\ M_\odot \leq M_{HB} \leq 0.55\ M_\odot}$ \\ \hline
   \multirow{2}{*}{M13}  & inner & 2 & 5 & 3 & 5  \\
   & outer & 4 & 2 & 7 & 3 \\
   \hline
   \multirow{2}{*}{M3}  & inner & - & - & - & -  \\
   & outer & 1 & 1 & - & 1 \\
   \hline
    \end{tabular} }
\end{table*}

By adopting the BaSTI-IAC stellar model library, we retrieved the post-ZAHB evolutionary tracks for various \Mhb, ${\rm [Fe/H]}=-1.55$ dex, ${\rm [\alpha/Fe]}=0.4$ dex, and Y $=$ 0.300. The evolutionary tracks for \Mhb\ $=$ 0.485 \Msun, 0.49 \Msun, 0.50 \Msun, 0.52 \Msun, 0.55 \Msun, and 0.60 \Msun\  along with TAHB and ZAHB loci are plotted in \autoref{fig:cmd_evol}. It is clear from the figure that the evolutionary track of ${\rm M_{HB}} = 0.55\ {\rm M_\odot}$ (violet solid line) is lying just above the TAHB locus (green dashed line) and the evolutionary track of ${\rm M_{HB}} = 0.60\ {\rm M_\odot}$ (violet dashed line) lying below the TAHB locus. This suggests that we can identify only those post-HB stars which have evolved from ${\rm M_{HB}} \leq 0.55\ {\rm M_\odot}$ and post-HB stars evolved from higher \Mhb\ would be indistinguishable from HB stars in the UV-optical CMDs. As proposed by \citet{Pietrinferni2006} and \cite{Chen2021}, stars with  ${\rm M_{HB}} \leq 0.55\ {\rm M_\odot}$ will never reach the tip of AGB (and hence do not experience any thermal pulse or the third dredge up) and will further evolve through the PEAGB phase. Hence, the observed post-HB stars in \autoref{fig:cmd_evol} will evolve through  either PEAGB or AGB-manqu\'e phase. We confirm that there are 31 post-HB stars in M13 and 3 post-HB stars in M3 which have evolved from the low mass HB stars (i.e., eHB and iHB stars, ${\rm M_{HB}} \leq 0.55\ {\rm M_\odot}$). In \autoref{tab:pHB}, we list the number of post-HB stars that evolved from the ZAHB locus having different HB mass ranges.

\subsection{SED of  HB and post-HB stars} \label{sec:sed_fitting}

\begin{figure*}
    \centering
    \includegraphics[width=0.49\textwidth]{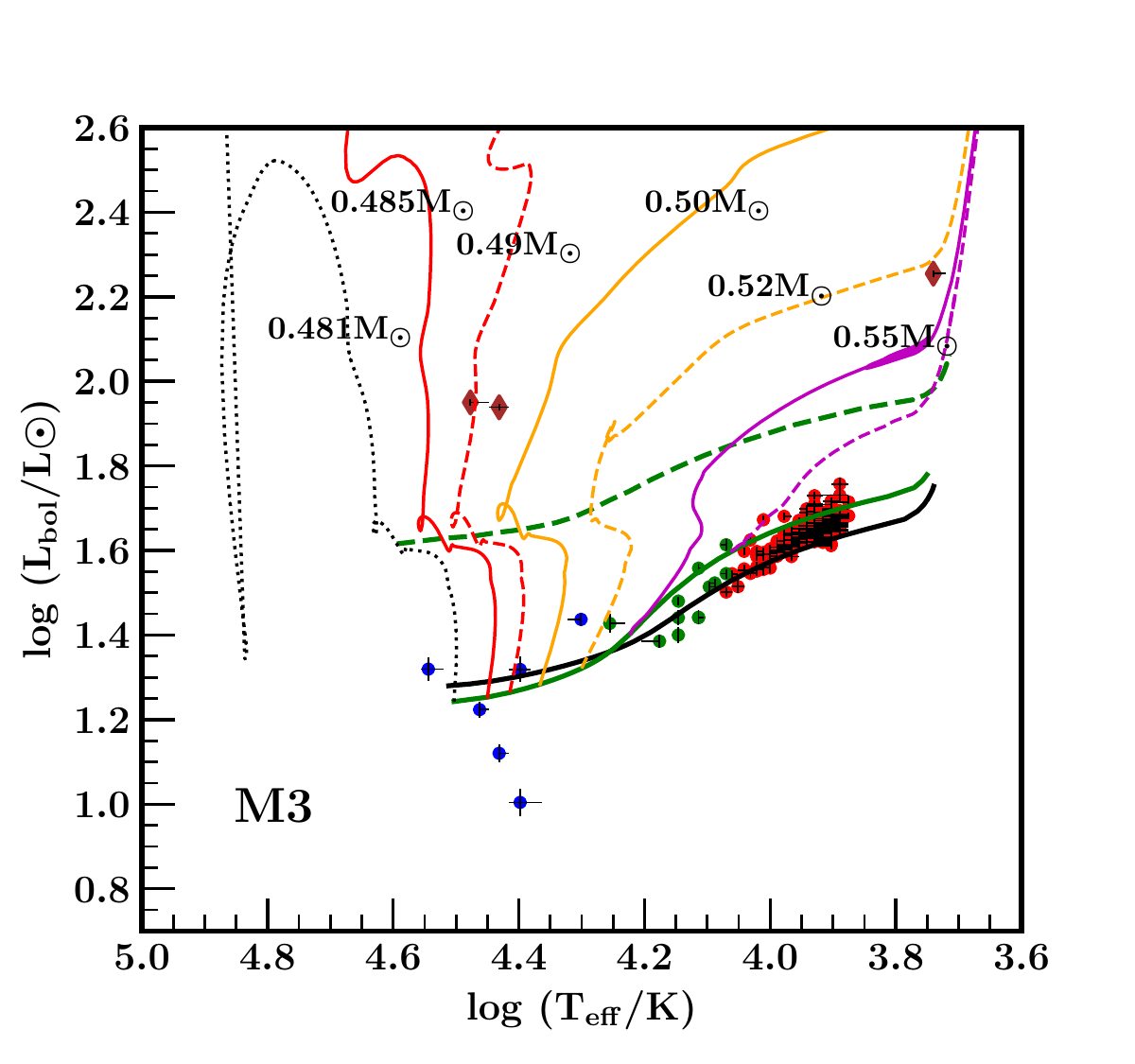}
    \includegraphics[width=0.49\textwidth]{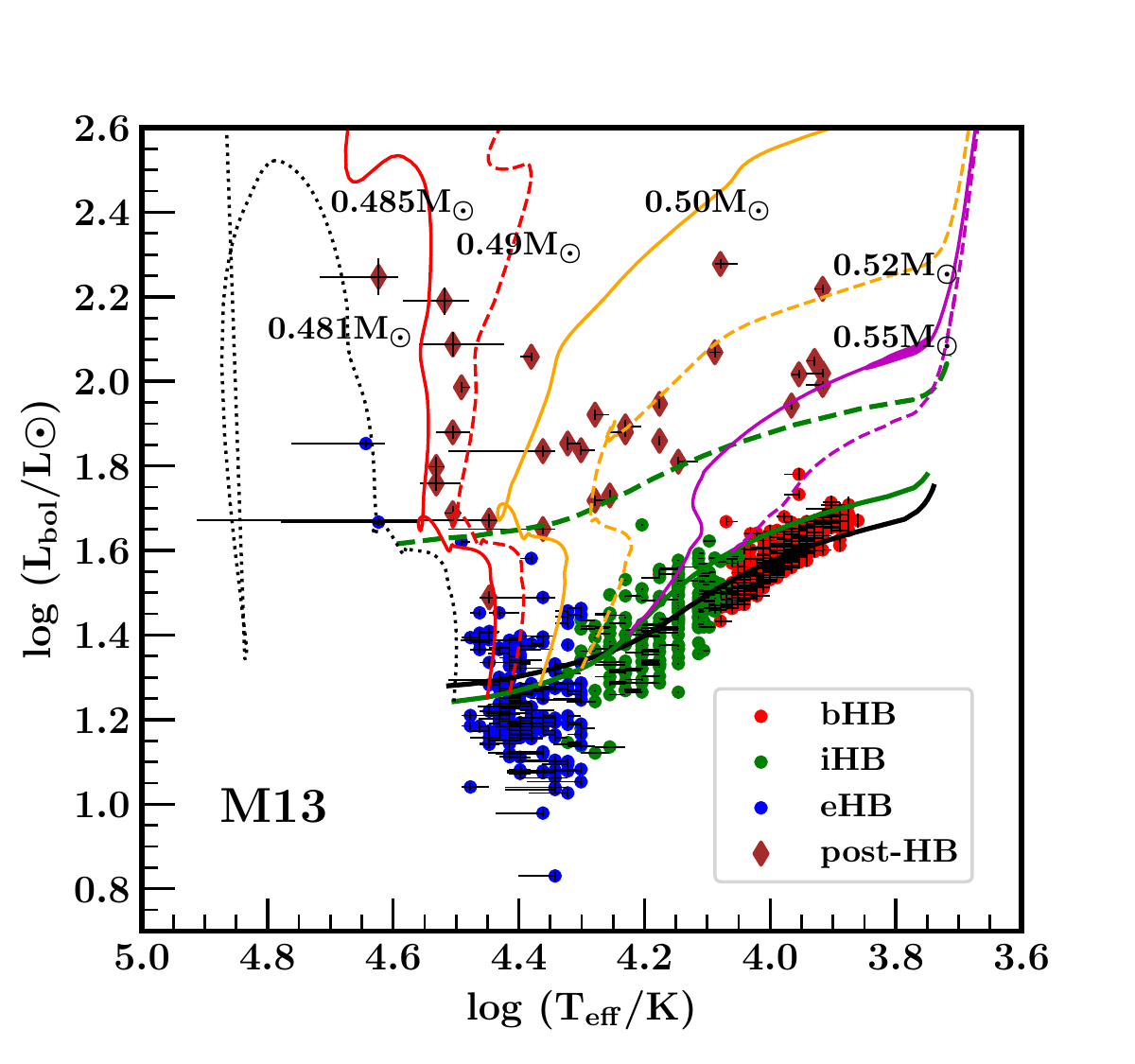}
    \caption{\Teff\ vs luminosity of HBs and post-HBs of M3 (left panel) and M13 (right panel). Various post-ZAHB evolutionary tracks of HB stars in the mass range $0.48-0.55$ \Msun, ZAHB and TAHB of ${\rm Y=0.300}$, and ZAHB of ${\rm Y=0.248}$ are shown in similar colours and styles as that of \autoref{fig:cmd_evol}. The post-ZAHB evolutionary track for \Mhb\ $=0.481$ \Msun\ is shown in a black dotted line. }
    \label{fig:Teff_lum_hb}
\end{figure*}

\begin{figure*}
    \centering
    \includegraphics[width=0.49\textwidth]{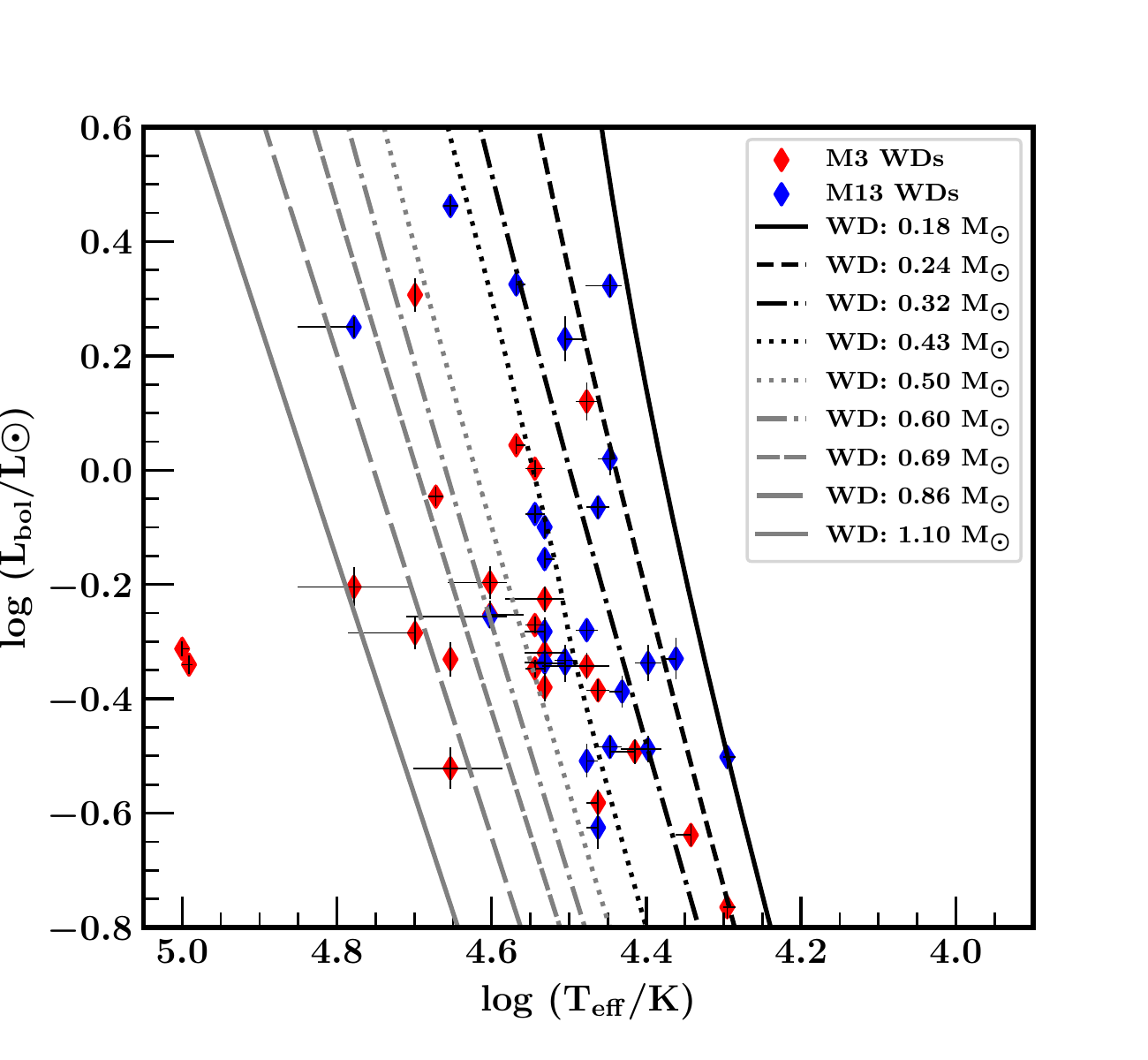}
    \includegraphics[width=0.49\textwidth]{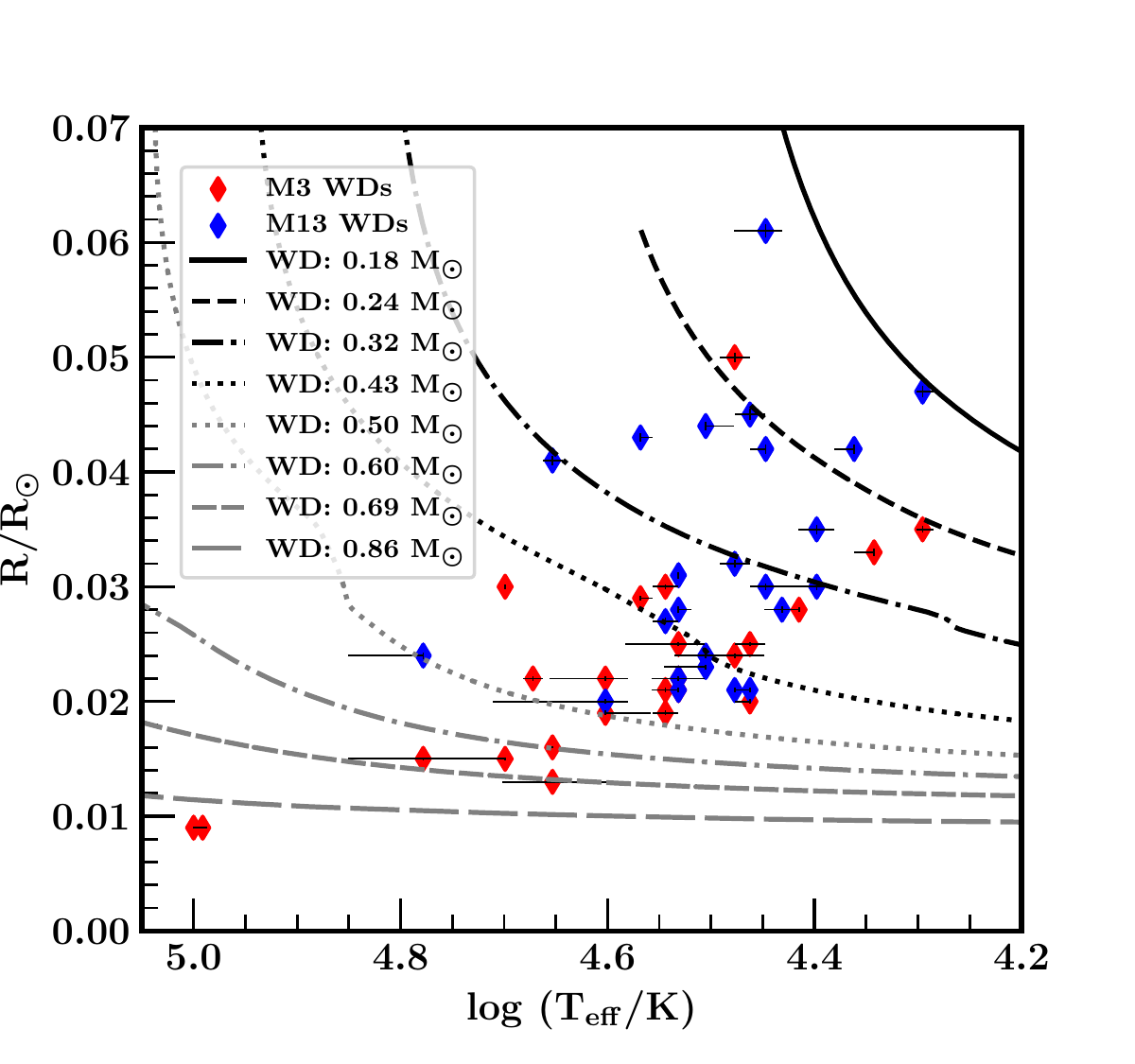}
    \caption{\Teff\ vs luminosity (left panel) and \Teff\ vs radius (right panel) of WDs of M3 (red diamonds) and M13 (blue diamonds). The cooling sequences of WDs in the mass range $0.18-1.10\ {\rm M_\odot}$ are shown in different styles and colours as mentioned in the legends. }
    \label{fig:Teff_lum_wd}
\end{figure*}

We performed spectral energy distribution (SED) fitting on HB and post-HB stars of M3 and M13 using observed photometric fluxes in UVIT and {\em HST} (UBVRI) filters for the inner (outer) region of the clusters. The SED fitting was performed in VO SED Analyzer\footnote{\href{http://svo2.cab.inta-csic.es/theory/vosa/}{http://svo2.cab.inta-csic.es/theory/vosa/}} \citep[][hearafter VOSA]{Bayo2008} where we have used Kurucz stellar atmosphere model \citep[][]{Castelli1997, Castelli2003} for bHB, iHB, eHB, and post-HB stars, and Tubingen NLTE Model Atmosphere Package (TMAP) Grid2 (Grid for H$=$1.0, ${\rm \lambda_{max}}\ \sim$ 4.0 $\times\ 10^5$ \AA, 20,000 K $<\ {\rm T_{eff}}\ <$ 150,000 K) model \citep{Werner1999J, Werner2003, Rauch2003} for eHB and post-HB stars. The best-fitted model parameters including the radial distance from the cluster center for the HB and post-HB stars are provided in Tables \ref{tab: sed_hb}, \ref{tab: sed_phb:M3}, and \ref{tab: sed_phb:M13}.  

In \autoref{fig:Teff_lum_hb}, we show the \Teff\ vs \Lbol\ plots of HB and post-HB stars along with BaSTI-IAC ZAHB locus for ${\rm Y=0.247}$ and $0.300$, and post-ZAHB evolutionary tracks for various HB masses (similar to those shown in \autoref{fig:cmd_evol}). We find 2 eHB stars of M13 (ID: 73 and 600) lying well above the TAHB locus (\Lbol\ $\sim$ 1.65 \Lsun\ and 1.85 \Lsun) which suggests that they have already evolved from their HB phase to post-HB phase. We have added them in the list of post-HB stars of M13 (\autoref{tab: sed_phb:M13}). We also see 2 more eHB stars (ID: 453 and 713) of M13 lying close to the TAHB locus (\Lbol\ $\sim$ 1.55 \Lsun\ and 1.60 \Lsun) which indicates that they might have almost completed their He-core burning stage and are about to evolve from their HB phase.

Out of three post-HB stars of M3, two are near the post-HB evolutionary track for \Mhb\ of 0.49 \Msun\ and the other one is near the post-HB evolutionary track of \Mhb\ of 0.52 \Msun\ (brown solids in the left panel of \autoref{fig:Teff_lum_hb}). We found that the two AGB-manqu\'e stars identified by \citet{Schiavon2012} in M3 are evolved from \Mhb\ of 0.49 \Msun. The post-HB stars of M13 can be found between post-HB evolutionary tracks of \Mhb\ of 0.481 \Msun\ and 0.55 \Msun\ (right panel of \autoref{fig:Teff_lum_hb}). We also found the \Lbol\ of one post-HB star (ID: 272) below the TAHB locus (\Lbol\ $\sim$ 1.45 \Lsun), so we excluded this star from the list of post-HB stars. 

\section{WDs of M3 and M13} \label{sec:wd}

Using UVIT observations, \citet{sneha2020} provided a catalogue of 107 WD candidates detected in 11 GCs. In FUV$-$optical CMDs, they lie below the eHB in a vertical sequence. We have considered the sources observed in similar regions of FUV$-$optical CMDs of M3 and M13 (\autoref{fig:uv_cmd}) as WDs candidates. We obtained a total of 27 WD candidates in M13 and 29 in M3. They show a narrow spread in F148W$-$F606W colour ($-3.2$ to $-1.4$ mag) in the inner region  whereas those observed in the outer region show a relatively larger spread ($-2.6$ to $1.0$ mag) in the F148W$-$V colour. 

We performed SED fitting on WD candidates of M3 and M13 using observed photometric fluxes in UVIT and {\em HST} (UBVRI) filters for the inner (outer) region of the clusters. The SED fitting was performed using two WD models, Koester WD models \citep{Koester2010}, and Levenhagen WD models \citep{Levenhagen2017}. Out of 56 WD candidates identified in M3 and M13, we could perform SED fitting only on 48 candidates using WD models (out of 48 WDs, 24 each are in M3 and M13). The SED-fitted stellar parameters of WD candidates of M3 and M13 are given in Tables \ref{tab: sed_wd:M3} and \ref{tab: sed_wd:M13}, respectively. Their \Teff, \Lbol, and radii are in the range of 19,750 $ - $ 100,000 K (19,750 $ - $ 60,000 K), 0.172 $ - $ 2.026 \Lsun\ (0.237 $ - $ 2.899 \Lsun), and 0.009 $ - $ 0.050 \Rsun\ (0.020 $ - $ 0.061 \Rsun), respectively. In \autoref{fig:Teff_lum_wd}, we show their \Teff\ vs. \Lbol\  (left panel) and \Teff\ vs. radius (right panel) for WD candidates of M3 and M13 in red and blue colours, respectively. We find WD candidates of M3 to be hotter and smaller in size when compared to those of M13 (median \Teff\ of M3 and M13 are 35,000 and 31,000 K, respectively; median WD radii of M3 and M13 are 0.022 and 0.03 \Rsun, respectively). 

We have also shown the WD cooling sequences in the mass-range $0.18-1.10$ \Msun. The cooling sequences of extremely low mass (ELM) He-core WDs (0.15 $\leq$ M/\Msun $\leq$ 0.43) are taken from \citet{Althaus2013} which are generated by computing the non-conservative evolution of a binary system consisting of an initially 1.0 \Msun\ zero-age main-sequence (ZAMS) star and a 1.4 \Msun\ neutron star for various initial orbital periods. The cooling tracks of  CO-core WDs were taken from \citet{Renedo2010} for WD mass range $0.50 - 0.86$ \Msun. These cooling tracks are derived from normal single stellar evolution. The cooling tracks for massive WDs ($1.06 - 1.28$ \Msun) were obtained from \citet{Althaus2007} which are computed under partial degenerate conditions of core carbon burning of WDs. 

The masses of WDs were estimated from the cooling tracks on the \Teff\ vs. \Lbol\ plot (left panel of \autoref{fig:Teff_lum_wd}). For candidates lying in between the cooling tracks, we assigned their masses as the mean mass of both the nearest cooling tracks. We have listed the estimated mass of WDs of M3 and M13 in Tables \ref{tab: sed_wd:M3} and \ref{tab: sed_wd:M13}. The WD candidates of M3 and M13 appear to have masses $~$0.18 \Msun\ or higher. There are 7 WD candidates of M3 and 12 WD candidates of M13 lying within the ELM WD cooling sequence (${\rm M_{WD} \leq 0.43\ M_\odot}$) whereas 15 WD candidates of both M3 and M13 have masses in the range, $0.43 - 0.50$ \Msun\ such that they could be formed through single or binary evolution. In M3, 5 WD candidates have masses of more than 0.86 \Msun. Out of them, 2 are with ${\rm M_{WD} \sim}$ 0.86 \Msun\ and one with ${\rm M_{WD} \sim 1.1\ M_\odot}$ which burn carbon in partially degenerate conditions. The rest 2 WD candidates having ${\rm \log (T_{eff}/K) \sim 5.0}$ are hotter than any WD cooling sequence to be fitted, hence, we could not predict their evolutionary status. 

Although \citet{Chen2021} studied the WDs of M3 and M13, their study was only focused on the WDs with masses 0.5$-$0.6 \Msun\ (single star evolution). In this paper, we see that WDs with such masses are less in number (\citet{Chen2021} and also suggested that the WD counts would be very less in number at higher luminosities. However, WDs evolved from the binary evolution would stay longer at higher luminosities and will have low masses than the single stellar evolution WDs \citep{Althaus2013}. This scenario may be supported by the fact that we see more than 53\% of HB stars lying below the ZAHB locus (\autoref{fig:Teff_lum_hb}) which may be blue-hook stars \citep[a result of hot flasher scenario from the binary system at RGB phase;][]{Lei2015, Lei2016} and as a result, they may end up with low mass He-core WDs which is the case we see in M13 compared to M3. It is also possible that the WDs from single-star evolution could cool faster than the low-mass WDs, and therefore become fainter rapidly beyond the detection limit of UVIT. This may also cause a reduction in the detected number of these WDs. In either case, this study presents the detection of low-mass WD candidates in M3 and M13. 

\section{Summary} \label{sec:conclusion}

Below we summarise the results and conclusions of this study:

\begin{itemize}
    \item We have studied the HB, post-HB, and hot WDs of M3 and M13 GCs using UVIT observations in two FUV filters in combination with the {\em HST} WFC3 filters for the inner region and the ground-based UBVRI photometry for the outer region of the cluster.
    \item We detected a total of 663 and 217 HB stars in UVIT filters for M13 and M3, respectively. The HB stars were further divided into three sub-populations, bHB, iHB, and eHB, based upon their \Teff\ in the range $7.75-11.5$ kK, $11.5-20$ kK, and $20-30$ kK, respectively. 
    \item We used synthetic colours of HB stars to find the He-spread for bHB, iHB, and eHB sub-populations of M13 as $0.247-0.256$, $0.26-0.30$, and $0.29-0.31$, respectively. The bHB of M3 has a He-spread of $0.252-0.266$. We found 3 and 31 post-HB stars in M3 and M13, respectively. Their evolutionary tracks on CMDs as well as \Teff\ vs. \Lbol\ plane suggest that they have evolved from \Mhb\ $\leq 0.55$ \Msun. 
    \item The comparison of synthetic and observed colours of HB stars suggests that the mass-loss at RGB and initial He-spread in both clusters have a simultaneous effect on the different HB distributions observed in M3 and M13. e.g., HB stars of M13 require a larger spread in Y as well as ${\rm \Delta M}$. M13 with a larger fraction of eHB and iHB stars than M3 can only be simulated if HB distribution of M13 has higher values of Y and ${\rm \Delta M}$. 

    \item We compared the derived He-abundances and RGB mass-loss with the latest available estimates obtained using multiple populations analysis by \cite{Milone2018, Tailo2019, Tailo2020} and for the clusters, the values match well within the present  uncertainties/indetermination. We found that the observed blue HB stars of M13 are the progeny of 1G, 2G, and 2Ge RGB populations whereas the observed blue HB stars of M3 are the progeny of 2G RGB populations.
    
    \item We detected 24 WD candidates each in M3 and M13. Their parameters have a large range (\Teff, \Lbol, and stellar radii in the range 19,750$-$100,000 K (19,750$-$60,000 K), 0.172$-$2.026 \Lsun\ (0.237$-$2.899 \Lsun), 0.009$-$0.050 \Rsun\ (0.020$-$0.061 \Rsun), respectively, for M3 (M13)).
    \item The WD cooling sequences obtained from single, binary, and massive star evolution when over-plotted on the \Teff\ vs. \Lbol\ and \Teff\ vs. stellar radii diagram suggest that some of the FUV detected WD candidates of M13 could be evolved from binary stellar evolution (lying on ELM cooling sequences), whereas most of the FUV detected WD candidates of M3 follow single stellar evolution.
\end{itemize}   

\section*{Acknowledgements}
We thank the referee for his/her suggestions which improved the content and readability of our manuscript. RK would like to acknowledge CSIR Research Fellowship (JRF) Grant No. 09/983(0034)/2019-EMR-1 for financial support. ACP 
and SP acknowledge the support of Indian Space Research Organisation (ISRO) under \mbox{{\em AstroSat}} archival Data utilization program. This publication uses the data from the \mbox{{\em AstroSat}} mission of the ISRO, archived at the Indian Space Science Data Center (ISSDC). AS acknowledges support from SERB Power Fellowship. DKO acknowledges the support of the Department of Atomic Energy, Government of India, under Project Identification No. RTI 4002. ACP also thanks Inter University centre for Astronomy and Astrophysics (IUCAA), Pune, India for providing facilities under associateship programs.

\section*{DATA AVAILABILITY}
The data underlying this article will be shared upon reasonable request to the corresponding author.




\bibliographystyle{mnras}
\bibliography{draft}


\begin{table*}
\caption{Stellar parameters along with their associated errors of 217 HBs of M3 and 678 HBs of M13 derived from SED fitting. The eTeffp and eTeffm columns are the upper and lower limits of error in \Teff. The complete table will be available online.}
\label{tab: sed_hb}
 \adjustbox{max width=\textwidth}{
\begin{tabular}{|r|r|r|r|r|r|r|r|r|r|r|r|r|r|l|c|}
\hline
  \multicolumn{1}{|c|}{Cluster} &
  \multicolumn{1}{|c|}{Phase} &
  \multicolumn{1}{|c|}{ID} &
  \multicolumn{1}{c|}{RA} &
  \multicolumn{1}{c|}{DEC} &
  \multicolumn{1}{c|}{Teff} &
  \multicolumn{1}{c|}{eTeffp} &
  \multicolumn{1}{c|}{eTeffm} &
  \multicolumn{1}{c|}{logg} &
  \multicolumn{1}{c|}{elogg} &
  \multicolumn{1}{c|}{L} &
  \multicolumn{1}{c|}{eL} &
  \multicolumn{1}{c|}{R} &
  \multicolumn{1}{c|}{eR} &
  \multicolumn{1}{c|}{Model} &
  \multicolumn{1}{c|}{Radial distance} \\
  
        & & &
  \multicolumn{1}{c|}{(degree)} &
  \multicolumn{1}{c|}{(degree)} &
  \multicolumn{1}{c|}{(K)} &
  \multicolumn{1}{c|}{(K)} &
  \multicolumn{1}{c|}{(K)} &
  \multicolumn{1}{c|}{(dex)} &
  \multicolumn{1}{c|}{(dex)} &
  \multicolumn{1}{c|}{(\Lsun)} &
  \multicolumn{1}{c|}{(\Lsun)} &
  \multicolumn{1}{c|}{(\Rsun)} &
  \multicolumn{1}{c|}{(\Rsun)} &
  \multicolumn{1}{l|}{} &
  \multicolumn{1}{c|}{(arcsec)} \\
  
\hline
 M3 & bHB & 27 & 205.44031 & 28.29328 & 9250.0 & 250.0 & 0.0 & 2.7 & 0.51 & 39.433 & 0.917 & 2.446 & 0.0197 & Kurucz2003alp & 456.9\\
 M3 & bHB &  28 & 205.4633 & 28.2767 & 7750.0 & 500.0 & 0.0 & 3.15 & 0.84 & 45.032 & 1.577 & 3.718 & 0.03   & Kurucz2003 & 451.5\\
 M3 & bHB &  34 & 205.36681 & 28.40855 & 8000.0 & 250.0 & 0.0 & 1.9 & 1.04 & 41.055 & 0.886 & 3.335 & 0.0269 & Kurucz2003alp & 586.1\\
 M3 & bHB &  63 & 205.4456 & 28.39114 & 8500.0 & 250.0 & 250.0 & 2.8 & 1.03 & 43.934 & 1.761 & 3.055 & 0.0246 & Kurucz2003alp & 329.4\\
 M3 & bHB &  68 & 205.4977 & 28.33659 & 8000.0 & 250.0 & 0.0 & 2.55 & 1.25 & 42.871 & 0.93 & 3.415 & 0.0275 & Kurucz2003 & 217.4\\
 \hline\end{tabular}
}
\end{table*}

\begin{table*}
\caption{Stellar parameters along with their associated errors of 3 post-HBs of M3 derived from SED fitting.}
\label{tab: sed_phb:M3}
 \adjustbox{max width=\textwidth}{
\begin{tabular}{|r|r|r|r|r|r|r|r|r|r|r|r|l|c|}
\hline
  \multicolumn{1}{|c|}{ID} &
  \multicolumn{1}{c|}{RA} &
  \multicolumn{1}{c|}{DEC} &
  \multicolumn{1}{c|}{Teff} &
  \multicolumn{1}{c|}{eTeffp} &
  \multicolumn{1}{c|}{eTeffm} &
  \multicolumn{1}{c|}{logg} &
  \multicolumn{1}{c|}{elogg} &
  \multicolumn{1}{c|}{L} &
  \multicolumn{1}{c|}{eL} &
  \multicolumn{1}{c|}{R} &
  \multicolumn{1}{c|}{eR} &

  \multicolumn{1}{c|}{Model} &
  \multicolumn{1}{c|}{Radial distance} \\
 
      &
  \multicolumn{1}{c|}{(degree)} &
  \multicolumn{1}{c|}{(degree)} &
  \multicolumn{1}{c|}{(K)} &
  \multicolumn{1}{c|}{(K)} &
  \multicolumn{1}{c|}{(K)} &
  \multicolumn{1}{c|}{(dex)} &
  \multicolumn{1}{c|}{(dex)} &
  \multicolumn{1}{c|}{(\Lsun)} &
  \multicolumn{1}{c|}{(\Lsun)} &
  \multicolumn{1}{c|}{(\Rsun)} &
  \multicolumn{1}{c|}{(\Rsun)} &

  \multicolumn{1}{c|}{} &
  \multicolumn{1}{c|}{(arcsec)} \\
  
\hline
  92 & 205.52431 & 28.31793 & 30000.0 & 0.0 & 2000.0 & 6.35 & 0.55 & 89.171 & 1.634 & 0.349 & 0.0028 & tmap1 & 226.9\\
  174 & 205.5049 & 28.39036 & 27000.0 & 1000.0 & 1000.0 & 6.1 & 0.62 & 86.836 & 1.62 & 0.425 & 0.0034 & tmap1 & 145.7\\
  196 & 205.5928 & 28.28496 & 5500.0 & 0.0 & 250.0 & 3.65 & 0.9 & 179.776 & 3.488 & 14.759 & 0.1189 & Kurucz2003alp & 360.9\\
\hline\end{tabular}
}
\end{table*}


\begin{table*}
\caption{Stellar parameters of 31 post-HBs of M13 derived from SED fitting. Star ID 73 and 600 are added in the list of post-HB stars on the basis of their position in the post-HB region in the log \Teff\ vs. log \Lbol\ plane. }
\label{tab: sed_phb:M13}
 \adjustbox{max width=\textwidth}{
\begin{tabular}{|r|r|r|r|r|r|r|r|r|r|r|r|l|c|}
\hline
  \multicolumn{1}{|c|}{ID} &
  \multicolumn{1}{c|}{RA} &
  \multicolumn{1}{c|}{DEC} &
  \multicolumn{1}{c|}{Teff} &
  \multicolumn{1}{c|}{eTeffp} &
  \multicolumn{1}{c|}{eTeffm} &
  \multicolumn{1}{c|}{logg} &
  \multicolumn{1}{c|}{elogg} &
  \multicolumn{1}{c|}{L} &
  \multicolumn{1}{c|}{eL} &
  \multicolumn{1}{c|}{R} &
  \multicolumn{1}{c|}{eR} &

  \multicolumn{1}{c|}{Model} &

  \multicolumn{1}{c|}{Radial distance} \\
  
      &
  \multicolumn{1}{c|}{(degree)} &
  \multicolumn{1}{c|}{(degree)} &
  \multicolumn{1}{c|}{(K)} &
  \multicolumn{1}{c|}{(K)} &
  \multicolumn{1}{c|}{(K)} &
  \multicolumn{1}{c|}{(dex)} &
  \multicolumn{1}{c|}{(dex)} &
  \multicolumn{1}{c|}{(\Lsun)} &
  \multicolumn{1}{c|}{(\Lsun)} &
  \multicolumn{1}{c|}{(\Rsun)} &
  \multicolumn{1}{c|}{(\Rsun)} &

  \multicolumn{1}{c|}{} &

  \multicolumn{1}{c|}{(arcsec)} \\
\hline
  19 & 250.43491 & 36.36537 & 28000.0 & 30000.0 & 1000.0 & 5.5 & 2.29 & 46.889 & 1.817 & 0.29 & 0.003 & tmap2  & 342.3\\
  73 & 250.4579 & 36.39996 & 42000.0 & 15000.0 & 6000.0 & 4.05 & 0.15 & 46.462 & 0.994 & 0.128 & 0.0013 & tmap2  & 239.6 \\
  139 & 250.40021 & 36.4326 & 17000.0 & 1000.0 & 0.0 & 4.85 & 0.23 & 75.757 & 1.981 & 1.008 & 0.0103 &  Kurucz2003alp  & 116.4\\
  374 & 250.39349 & 36.45731 & 24000.0 & 1000.0 & 0.0 & 3.45 & 0.47 & 114.338 & 2.698 & 0.612 & 0.0063 &  Kurucz2003 &  82.6\\
  423 & 250.4183 & 36.45238 & 8250.0 & 500.0 & 0.0 & 2.1 & 0.62 & 98.036 & 2.175 & 4.806 & 0.0492 & Kurucz2003alp &  28.8\\
  428 & 250.3956 & 36.45974 & 34000.0 & 1000.0 & 1000.0 & 4.0 & 0.0 & 62.785 & 3.367 & 0.228 & 0.0023 &  Kurucz2003alp &  76.0\\
  441 & 250.39751 & 36.45974 & 34000.0 & 2000.0 & 3000.0 & 3.9 & 0.2 & 57.31 & 2.14 & 0.219 & 0.0022 &  Kurucz2003 &  70.4\\
  492 & 250.4493 & 36.44659 & 32000.0 & 1000.0 & 1000.0 & 4.2 & 0.4 & 48.775 & 1.789 & 0.227 & 0.0023 &  Kurucz2003alp &  92.8\\
  521 & 250.4314 & 36.45312 & 9250.0 & 250.0 & 0.0 & 3.25 & 0.78 & 87.663 & 1.984 & 3.649 & 0.0374 & Kurucz2003 &  36.8\\
  600 & 250.3828 & 36.47147 & 44000.0 & 12000.0 & 3000.0 & 4.0 & 0.0 & 71.268 & 1.521 & 0.144 & 0.0015 &  tmap2  & 120.5 \\
  603 & 250.2832 & 36.50146 & 15000.0 & 0.0 & 0.0 & 3.8 & 0.9 & 88.492 & 2.262 & 1.396 & 0.0143 &  Kurucz2003  & 428.3\\
  575 & 250.44 & 36.45331 & 8250.0 & 250.0 & 0.0 & 1.7 & 0.64 & 165.451 & 3.722 & 6.297 & 0.0645 & Kurucz2003alp &  57.7\\
  620 & 250.4585 & 36.44939 & 17000.0 & 0.0 & 1000.0 & 4.8 & 0.33 & 78.27 & 2.921 & 1.019 & 0.0104 &  Kurucz2003alp  & 112.7\\
  671 & 250.4241 & 36.46192 & 18000.0 & 0.0 & 0.0 & 3.3 & 0.87 & 53.725 & 1.554 & 0.754 & 0.0077 &  Kurucz2003alp &  9.9\\
  684 & 250.4384 & 36.45827 & 21000.0 & 0.0 & 1000.0 & 3.75 & 0.78 & 71.279 & 2.726 & 0.639 & 0.0065 &  Kurucz2003alp &  48.3\\
  774 & 250.342 & 36.49117 & 19000.0 & 0.0 & 1000.0 & 3.75 & 0.84 & 52.243 & 1.458 & 0.672 & 0.0069 &  Kurucz2003alp  & 257.1\\
  753 & 250.4156 & 36.46846 & 9000.0 & 250.0 & 0.0 & 2.3 & 0.75 & 103.918 & 2.311 & 4.194 & 0.043 &  Kurucz2003 &  35.8\\
  794 & 250.4182 & 36.46939 & 8250.0 & 500.0 & 0.0 & 2.2 & 0.81 & 104.518 & 2.474 & 4.993 & 0.0511 &  Kurucz2003alp &  35.9\\
  929 & 250.35921 & 36.49708 & 19000.0 & 0.0 & 1000.0 & 4.1 & 0.92 & 83.429 & 2.497 & 0.847 & 0.0087 &  Kurucz2003alp  & 225.4\\
  875 & 250.43629 & 36.46927 & 31000.0 & 0.0 & 1000.0 & 3.55 & 0.15 & 96.84 & 2.674 & 0.341 & 0.0035 &  Kurucz2003alp &  53.8\\
  941 & 250.39461 & 36.48714 & 12000.0 & 250.0 & 750.0 & 5.0 & 0.0 & 189.535 & 4.853 & 3.194 & 0.0327 &  Kurucz2003alp  & 125.9\\
  947 & 250.411 & 36.48284 & 20000.0 & 0.0 & 1000.0 & 3.7 & 1.08 & 68.718 & 1.848 & 0.697 & 0.0071 &  Kurucz2003alp  & 88.5\\
  955 & 250.4122 & 36.48304 & 32000.0 & 2000.0 & 2000.0 & 3.75 & 0.25 & 75.788 & 2.861 & 0.284 & 0.0029 &  Kurucz2003alp &  88.0\\
  958 & 250.42909 & 36.47835 & 12250.0 & 250.0 & 250.0 & 3.25 & 0.84 & 117.174 & 3.109 & 2.406 & 0.0246 &  Kurucz2003alp &  69.8\\
  974 & 250.39751 & 36.48886 & 14000.0 & 0.0 & 1000.0 & 4.05 & 0.76 & 64.549 & 1.662 & 1.364 & 0.014 &  Kurucz2003alp  & 125.9\\
  1068 & 250.45441 & 36.48516 & 23000.0 & 8000.0 & 1000.0 & 6.45 & 2.45 & 68.387 & 2.183 & 0.518 & 0.0053 &  tmap2  & 131.1\\
  1115 & 250.4016 & 36.51425 & 8500.0 & 250.0 & 0.0 & 2.0 & 0.5 & 111.766 & 2.826 & 4.849 & 0.0497 &   Kurucz2003alp  & 204.4\\
  1147 & 250.4494 & 36.52553 & 33000.0 & 5000.0 & 3000.0 & 4.05 & 0.15 & 154.973 & 11.45 & 0.378 & 0.0039 &  tmap2  & 249.5\\
  1156 & 250.49001 & 36.51987 & 15000.0 & 0.0 & 0.0 & 4.4 & 0.49 & 72.32 & 1.78 & 1.268 & 0.013 & Kurucz2003alp  & 292.6\\
  1164 & 250.2112 & 36.61396 & 42000.0 & 9000.0 & 3000.0 & 4.45 & 0.15 & 176.72 & 17.836 & 0.249 & 0.0026 & tmap2  & 824.0\\
  1180 & 250.5022 & 36.5579 & 32000.0 & 4000.0 & 6000.0 & 4.6 & 1.48 & 122.214 & 8.07 & 0.357 & 0.0037 & tmap2  & 422.7\\
  
\hline\end{tabular}
}
\end{table*}


\begin{table*}
\caption{Stellar parameters along with the associated errors of 24 WDs of M3 derived from SED fitting. The eMassp and eMassm columns are the upper and lower limits of error in WD mass.}
\label{tab: sed_wd:M3}
 \adjustbox{max width=\textwidth}{
\begin{tabular}{|r|r|r|r|r|r|r|r|r|r|r|r|l|c|r|r|r|}
\hline
  \multicolumn{1}{|c|}{ID} &
  \multicolumn{1}{c|}{RA} &
  \multicolumn{1}{c|}{DEC} &
  \multicolumn{1}{c|}{Teff} &
  \multicolumn{1}{c|}{eTeffp} &
  \multicolumn{1}{c|}{eTeffm} &
  \multicolumn{1}{c|}{logg} &
  \multicolumn{1}{c|}{elogg} &
  \multicolumn{1}{c|}{L} &
  \multicolumn{1}{c|}{eL} &
  \multicolumn{1}{c|}{R} &
  \multicolumn{1}{c|}{eR} &

  \multicolumn{1}{c|}{Model} &

  \multicolumn{1}{c|}{Radial distance} &
  \multicolumn{1}{c|}{Mass} &
  \multicolumn{1}{c|}{eMassp} &
  \multicolumn{1}{c|}{eMassm} \\
  
    &
  \multicolumn{1}{c|}{(degree)} &
  \multicolumn{1}{c|}{(degree)} &
  \multicolumn{1}{c|}{(K)} &
  \multicolumn{1}{c|}{(K)} &
  \multicolumn{1}{c|}{(K)} &
  \multicolumn{1}{c|}{(dex)} &
  \multicolumn{1}{c|}{(dex)} &
  \multicolumn{1}{c|}{(\Lsun)} &
  \multicolumn{1}{c|}{(\Lsun)} &
  \multicolumn{1}{c|}{(\Rsun)} &
  \multicolumn{1}{c|}{(\Rsun)} &

  \multicolumn{1}{c|}{} &

  \multicolumn{1}{c|}{(arcsec)} &
  \multicolumn{1}{c|}{(\Msun)} &
  \multicolumn{1}{c|}{(\Rsun)} &
  \multicolumn{1}{c|}{(\Rsun)} \\
 
\hline
  57 & 205.4805 & 28.33584 & 19750.0 & 250.0 & 500.0 & 9.3 & 0.19 & 0.172 & 0.008 & 0.035 & 3.0E-4 & koester2  & 261.8 & 0.24 & 0.01 & 0.01\\
  71 & 205.4675 & 28.37843 & 22000.0 & 1000.0 & 0.0 & 8.32 & 0.45 & 0.23 & 0.01 & 0.033 & 3.0E-4 & koester2  & 256.3 & 0.28 & 0.04 & 0.04\\
  223 & 205.5278 & 28.37083 & 45000.0 & 0.0 & 0.0 & 7.62 & 0.72 & 0.467 & 0.032 & 0.016 & 1.0E-4 & koester2 &  69.3 & 0.78 & 0.08 & 0.08\\
  232 & 205.52299 & 28.37996 & 29000.0 & 1000.0 & 0.0 & 8.82 & 0.45 & 0.262 & 0.014 & 0.02 & 2.0E-4 & koester2 &  81.1 & 0.47 & 0.035 & 0.01\\
  266 & 205.53951 & 28.36581 & 30000.0 & 1000.0 & 1000.0 & 9.34 & 0.14 & 1.319 & 0.102 & 0.05 & 4.0E-4 & levenhagen17 &  50.0 & 0.28 & 0.04 & 0.04\\
  291 & 205.52631 & 28.38557 & 34000.0 & 4000.0 & 2000.0 & 9.32 & 0.2 & 0.595 & 0.031 & 0.025 & 2.0E-4 & koester2 &  76.1 & 0.47 & 0.035 & 0.035\\
  352 & 205.54311 & 28.37141 & 100000.0 & 0.0 & 2000.0 & 7.53 & 0.14 & 0.487 & 0.014 & 0.009 & 1.0E-4 & levenhagen17 &  27.0 & 99.0 & 99.0 & 99.0\\
  361 & 205.5479 & 28.36636 & 98000.0 & 2000.0 & 1000.0 & 7.32 & 0.28 & 0.457 & 0.017 & 0.009 & 1.0E-4 & levenhagen17 &  39.3 & 99.0 & 99.0 & 99.0\\
  474 & 205.5365 & 28.38878 & 47000.0 & 1000.0 & 1000.0 & 9.37 & 0.11 & 0.899 & 0.035 & 0.022 & 2.0E-4 & levenhagen17 &  56.0 & 0.65 & 0.05 & 0.05\\
  619 & 205.54179 & 28.39195 & 29000.0 & 1000.0 & 1000.0 & 9.18 & 0.28 & 0.412 & 0.019 & 0.025 & 2.0E-4 & koester2 &  56.8 & 0.395 & 0.035 & 0.035\\
  652 & 205.55 & 28.38515 & 60000.0 & 10000.0 & 10000.0 & 6.9 & 0.34 & 0.625 & 0.049 & 0.015 & 1.0E-4 & koester2 &  28.8 & 1.1 & 0.24 & 0.24\\
  661 & 205.56799 & 28.36277 & 50000.0 & 10000.0 & 0.0 & 7.0 & 0.35 & 0.519 & 0.034 & 0.015 & 1.0E-4 & koester2 &  81.1 & 0.86 & 0.24 & 0.01\\
  672 & 205.5647 & 28.36751 & 45000.0 & 5000.0 & 7000.0 & 9.25 & 0.25 & 0.301 & 0.025 & 0.013 & 1.0E-4 & koester2 &  62.4 & 0.98 & 0.12 & 0.12\\
  761 & 205.5547 & 28.38568 & 35000.0 & 1000.0 & 1000.0 & 7.45 & 0.31 & 0.536 & 0.018 & 0.021 & 2.0E-4 & koester2 &  36.2 & 0.47 & 0.035 & 0.035\\
  798 & 205.56129 & 28.38363 & 30000.0 & 4000.0 & 2000.0 & 9.32 & 0.2 & 0.454 & 0.024 & 0.024 & 2.0E-4 & koester2 &  46.7 & 0.395 & 0.035 & 0.035\\
  811 & 205.5656 & 28.37999 & 34000.0 & 2000.0 & 2000.0 & 9.25 & 0.25 & 0.479 & 0.025 & 0.022 & 2.0E-4 & koester2 &  55.3 & 0.47 & 0.035 & 0.035\\
  826 & 205.56329 & 28.38589 & 35000.0 & 1000.0 & 1000.0 & 6.82 & 0.28 & 0.45 & 0.016 & 0.019 & 2.0E-4 & koester2 &  56.4 & 0.5 & 0.01 & 0.01\\
  844 & 205.5614 & 28.39203 & 40000.0 & 0.0 & 4000.0 & 6.9 & 0.34 & 0.559 & 0.031 & 0.019 & 2.0E-4 & koester2 &  67.2 & 0.6 & 0.01 & 0.01\\
  868 & 205.5081 & 28.46718 & 35000.0 & 1000.0 & 1000.0 & 8.18 & 0.65 & 1.006 & 0.035 & 0.03 & 2.0E-4 & koester2  & 347.9 & 0.395 & 0.035 & 0.035\\
  1050 & 205.6338 & 28.42872 & 37000.0 & 0.0 & 1000.0 & 7.21 & 0.16 & 1.106 & 0.038 & 0.029 & 2.0E-4 & levenhagen17  & 327.7 & 0.44 & 0.035 & 0.035\\
  1055 & 205.5993 & 28.48572 & 26000.0 & 2000.0 & 0.0 & 9.1 & 0.3 & 0.322 & 0.016 & 0.028 & 2.0E-4 & koester2  & 422.3 & 0.34 & 0.02 & 0.02\\
  1112 & 205.5349 & 28.35661 & 50000.0 & 0.0 & 0.0 & 7.62 & 0.72 & 2.026 & 0.136 & 0.03 & 2.0E-4 & koester2 &  85.8 & 0.55 & 0.05 & 0.05\\
  1134 & 205.54939 & 28.39877 & 34000.0 & 0.0 & 0.0 & 8.75 & 0.29 & 0.417 & 0.024 & 0.021 & 2.0E-4 & levenhagen17 &  77.4 & 0.5 & 0.01 & 0.01\\
  1135 & 205.55901 & 28.39784 & 40000.0 & 5000.0 & 2000.0 & 9.15 & 0.32 & 0.636 & 0.042 & 0.022 & 2.0E-4 & koester2 &  81.3 & 0.55 & 0.05 & 0.05\\
\hline\end{tabular} 
}
\end{table*}

\begin{table*}
\caption{Stellar parameters of 24 WDs of M13 derived from SED fitting.}
\label{tab: sed_wd:M13}
 \adjustbox{max width=\textwidth}{
\begin{tabular}{|r|r|r|r|r|r|r|r|r|r|r|r|l|c|r|r|r|}
\hline
  \multicolumn{1}{|c|}{ID} &
  \multicolumn{1}{c|}{RA} &
  \multicolumn{1}{c|}{DEC} &
  \multicolumn{1}{c|}{Teff} &
  \multicolumn{1}{c|}{eTeffp} &
  \multicolumn{1}{c|}{eTeffm} &
  \multicolumn{1}{c|}{logg} &
  \multicolumn{1}{c|}{elogg} &
  \multicolumn{1}{c|}{L} &
  \multicolumn{1}{c|}{eL} &
  \multicolumn{1}{c|}{R} &
  \multicolumn{1}{c|}{eR} &
  \multicolumn{1}{c|}{Model} &

  \multicolumn{1}{c|}{Radial distance} &
  \multicolumn{1}{c|}{Mass} &
  \multicolumn{1}{c|}{eMassp} &
  \multicolumn{1}{c|}{eMassm} \\
  
      &
  \multicolumn{1}{c|}{(degree)} &
  \multicolumn{1}{c|}{(degree)} &
  \multicolumn{1}{c|}{(K)} &
  \multicolumn{1}{c|}{(K)} &
  \multicolumn{1}{c|}{(K)} &
  \multicolumn{1}{c|}{(dex)} &
  \multicolumn{1}{c|}{(dex)} &
  \multicolumn{1}{c|}{(\Lsun)} &
  \multicolumn{1}{c|}{(\Lsun)} &
  \multicolumn{1}{c|}{(\Rsun)} &
  \multicolumn{1}{c|}{(\Rsun)} &
  \multicolumn{1}{c|}{} &

  \multicolumn{1}{c|}{(arcsec)} &
  \multicolumn{1}{c|}{(\Msun)} &
  \multicolumn{1}{c|}{(\Rsun)} &
  \multicolumn{1}{c|}{(\Rsun)} \\
  
\hline
  6 & 250.4462 & 36.34667 & 29000.0 & 1000.0 & 0.0 & 7.24 & 0.2 & 0.237 & 0.02 & 0.021 & 2.0E-4 & levenhagen17  & 413.6 & 0.47 & 0.035 & 0.035\\
  36 & 250.3822 & 36.39391 & 19750.0 & 250.0 & 500.0 & 8.32 & 0.22 & 0.315 & 0.016 & 0.047 & 5.0E-4 & koester2  & 263.7 & 0.182 & 0.01 & 0.01\\
  45 & 250.33501 & 36.41498 & 34000.0 & 2000.0 & 0.0 & 6.82 & 0.28 & 0.522 & 0.029 & 0.022 & 2.0E-4 &  koester2  & 298.9 & 0.21 & 0.03 & 0.03\\
  46 & 250.25591 & 36.43906 & 27000.0 & 1000.0 & 0.0 & 8.6 & 0.49 & 0.41 & 0.026 & 0.028 & 3.0E-4 &  koester2  & 486.3 & 0.34 & 0.02 & 0.02\\
  74 & 250.5079 & 36.38501 & 32000.0 & 3000.0 & 0.0 & 6.85 & 0.32 & 0.459 & 0.034 & 0.023 & 2.0E-4 &  koester2  & 367.1 & 0.47 & 0.035 & 0.035\\
  149 & 250.26241 & 36.47466 & 25000.0 & 1000.0 & 1000.0 & 9.1 & 0.34 & 0.46 & 0.034 & 0.035 & 4.0E-4 &  koester2  & 464.6 & 0.28 & 0.04 & 0.04\\
  193 & 250.4158 & 36.43419 & 40000.0 & 10000.0 & 2000.0 & 9.25 & 0.25 & 0.555 & 0.022 & 0.02 & 2.0E-4 &  koester2 &  94.1 & 0.6 & 0.26 & 0.1\\
  287 & 250.4319 & 36.4391 & 45000.0 & 1000.0 & 1000.0 & 9.34 & 0.14 & 2.899 & 0.111 & 0.041 & 4.0E-4 &  levenhagen17 &  80.2 & 0.47 & 0.035 & 0.035\\
  289 & 250.40359 & 36.44783 & 34000.0 & 2000.0 & 0.0 & 9.2 & 0.24 & 0.461 & 0.019 & 0.021 & 2.0E-4 & koester2 &  68.3 & 0.47 & 0.035 & 0.035\\
  390 & 250.4189 & 36.45007 & 28000.0 & 1000.0 & 1000.0 & 9.37 & 0.11 & 0.328 & 0.015 & 0.03 & 3.0E-4 &  levenhagen17 &  36.3 & 0.395 & 0.035 & 0.035\\
  408 & 250.4214 & 36.45056 & 34000.0 & 0.0 & 1000.0 & 7.21 & 0.16 & 0.699 & 0.026 & 0.028 & 3.0E-4 &  levenhagen17 &  33.5 & 0.47 & 0.035 & 0.035\\
  413 & 250.3933 & 36.45935 & 32000.0 & 1000.0 & 1000.0 & 7.16 & 0.14 & 0.465 & 0.019 & 0.024 & 2.0E-4 &  levenhagen17 &  82.6 & 0.47 & 0.035 & 0.035\\
  434 & 250.4135 & 36.4547 & 34000.0 & 0.0 & 0.0 & 9.05 & 0.29 & 0.795 & 0.035 & 0.031 & 3.0E-4 & levenhagen17 &  30.5 & 0.43 & 0.01 & 0.01\\
  554 & 250.44569 & 36.45067 & 25000.0 & 2000.0 & 1000.0 & 9.2 & 0.24 & 0.325 & 0.017 & 0.03 & 3.0E-4 &  koester2 &  76.6 & 0.32 & 0.01 & 0.01\\
  687 & 250.41229 & 36.46652 & 60000.0 & 10000.0 & 0.0 & 7.22 & 0.6 & 1.78 & 0.072 & 0.024 & 3.0E-4 &  koester2 &  36.6 & 0.78 & 0.08 & 0.08\\
  701 & 250.5107 & 36.43714 & 30000.0 & 0.0 & 1000.0 & 7.0 & 0.35 & 0.31 & 0.021 & 0.021 & 2.0E-4 &  koester2  & 270.0 & 0.47 & 0.01 & 0.01\\
  712 & 250.4415 & 36.4586 & 35000.0 & 1000.0 & 1000.0 & 6.82 & 0.28 & 0.838 & 0.031 & 0.027 & 3.0E-4 & koester2 &  57.1 & 0.47 & 0.035 & 0.035\\
  760 & 250.4261 & 36.46552 & 37000.0 & 0.0 & 1000.0 & 9.21 & 0.24 & 2.114 & 0.09 & 0.043 & 4.0E-4 & levenhagen17 &  23.8 & 0.32 & 0.01 & 0.01\\
  772 & 250.4393 & 36.46181 & 28000.0 & 2000.0 & 1000.0 & 9.3 & 0.19 & 2.102 & 0.091 & 0.061 & 6.0E-4 & koester2 &  51.1 & 0.21 & 0.03 & 0.03\\
  860 & 250.43311 & 36.46903 & 29000.0 & 1000.0 & 1000.0 & 9.34 & 0.14 & 0.861 & 0.038 & 0.045 & 5.0E-4 & levenhagen17 &  46.4 & 0.28 & 0.04 & 0.04\\
  1051 & 250.43739 & 36.48792 & 30000.0 & 1000.0 & 1000.0 & 9.36 & 0.13 & 0.525 & 0.022 & 0.032 & 3.0E-4 & levenhagen17 &  110.6 & 0.36 & 0.01 & 0.01\\
  1160 & 250.3826 & 36.55949 & 28000.0 & 1000.0 & 0.0 & 7.03 & 0.39 & 1.047 & 0.071 & 0.042 & 4.0E-4 & koester2  & 376.2 & 0.24 & 0.01 & 0.01\\
  1166 & 250.5157 & 36.52699 & 32000.0 & 0.0 & 2000.0 & 8.68 & 0.52 & 1.696 & 0.155 & 0.044 & 4.0E-4 & koester2  & 363.6 & 0.28 & 0.04 & 0.04\\
  1184 & 250.52251 & 36.56129 & 23000.0 & 1000.0 & 0.0 & 8.52 & 0.55 & 0.468 & 0.039 & 0.042 & 4.0E-4 & koester2  & 467.1 & 0.21 & 0.03 & 0.03\\
\hline\end{tabular}
}
\end{table*}

\bsp	
\label{lastpage}
\end{document}